\documentclass[aps,reprint,superscriptaddress]{revtex4-1}
\usepackage{amssymb}
\usepackage{amsmath}
\usepackage{graphicx}
\usepackage{mathbbol}

\newcommand{\rmi}{\ensuremath{\mathrm{i}}}
\newcommand{\rme}{e}
\newcommand{\rmd}{d}

\newcommand{\pdiff}[2]{\frac{\partial\,#1}{\partial #2}}

\newcommand{\ket}[1]{\ensuremath{\left| #1\right\rangle}}
\newcommand{\braket}[2]{\ensuremath{\left\langle #1 | #2\right\rangle}}
\newcommand{\melement}[3]{\ensuremath{\left\langle #1\left| #2 \right| #3\right\rangle}}

\begin{document}

\title{Quantum beats in the polarization response of a dielectric to
  intense few-cycle laser pulses}

\author{M.~Korbman}
\affiliation{Max-Planck-Institut f\"ur Quantenoptik, Hans-Kopfermann-Stra{\ss}e 1, 85748 Garching, Germany}

\author{S.~ Yu.~Kruchinin}
\affiliation{Max-Planck-Institut f\"ur Quantenoptik, Hans-Kopfermann-Stra{\ss}e 1, 85748 Garching, Germany}

\author{V.~S.~Yakovlev}
\email{vladislav.yakovlev@lmu.de}
\affiliation{Ludwig-Maximilians-Universit\"at, Am Coulombwall 1, 85748 Garching, Germany}
\affiliation{Max-Planck-Institut f\"ur Quantenoptik, Hans-Kopfermann-Stra{\ss}e 1, 85748 Garching, Germany}

\begin{abstract}
  We have investigated the polarization response of a dielectric to
  intense few-cycle laser pulses with a focus on interband
  tunnelling. Once charge carriers are created in an initially empty
  conduction band, they make a significant contribution to the
  polarization response. In particular, the coherent superposition of
  conduction- and valence-band states results in quantum beats. This
  quantum-beat part of the polarization response
  is affected by the excitation dynamics and attosecond-scale motion
  of charge carriers in an intense laser field. Our analysis shows that, with the
  onset of Bloch oscillations or tunnelling, the nonlinear
  polarization response becomes sensitive to the carrier--envelope
  phase of the laser pulse.
\end{abstract}

\maketitle

\section{Introduction}
Multiphoton strong-field excitation of valence-band electrons has been intensively
investigated over several decades. Until recently, the main motivation
for experimental research has been laser damage of optical materials
\cite{Stuart_PRB_1996,Lenzner_PRL_1998,Mao_APA_2004}. Detailed time-resolved
measurements of electron dynamics used to be out of reach, as the
characteristic times of the most essential processes belong to the
attosecond domain. The situation has recently changed due to the
progress in the generation of few-cycle laser pulses and the
development of attosecond science
\cite{Hentschel_Nature_2001,Krausz_RMP_2009}. In this context, several
recent papers should be pointed out. Gertsvolf \emph{et al.}~\cite{Gertsvolf_JPB_2010} observed that an
elliptically polarized 40-fs pulse of near-infrared (NIR) radiation
changes its ellipticity as it propagates through a few micrometres of
fused silica. Based on their numerical
calculations, the authors concluded that the change of ellipticity is
due to sub-cycle, attosecond-scale excitation dynamics of valence-band
electrons. Their conclusions were confirmed by Mitrofanov \emph{et al.}~\cite{Mitrofanov_PRL_2011}, who
used a non-collinear pump--probe technique to observe an
attosecond-scale modulation of the electron density created by a 5-fs
NIR pulse in a $\mbox{SiO}_2$ sample. A
different optical strong-field phenomenon was observed by Ghimire
\emph{et al.}~\cite{Ghimire_Nature-Physics_2011} in ZnO exposed to high-power few-cycle mid-infrared laser
pulses: they observed the generation of high harmonics with properties
that cannot be explained by the conventional nonlinear optics. The origin of these harmonics was
linked to a combination of anharmonic electron motion and multiple
Bragg reflections at the boundaries of the first Brillouin zone, also
known as ``Bloch oscillations''~\cite{Ghimire_PRA_2012}. Another
manifestation of Bloch oscillations observed by the same team was a
redshift in the optical absorption in ZnO
\cite{Ghimire_PRL_2011}. Most recently, it was demonstrated that an
intense few-cycle laser pulse not only excites charge carriers in
$\mbox{SiO}_2$, but it also drives measurable currents that can be
steered by controlling the carrier--envelope phase (CEP)
\cite{Baltuska_Nature_2003} of the laser pulse
\cite{Schiffrin_Nature_2013}. The theoretical interpretation of that
experiment was given in terms of Wannier--Stark localization
\cite{Wannier_PR_1960,Durach_PRL_2011} that 
can be viewed as a frequency-domain description of
Bloch oscillations. These developments created a new perspective for
studying electron dynamics in solids exposed to strong fields, where
the emphasis is placed on attosecond-scale dynamics.

The foundations for the theoretical description of extremely nonlinear
interaction with strong fields, where conventional perturbation
theory breaks down, were laid by Keldysh in his seminal papers
\cite{Keldysh_JETP_1958,Keldysh_JETP_1965} where
multiphoton and tunnelling excitations~\cite{Wegener_2004},
which are two extreme regimes of interband transitions,
were distinguished by the Keldysh parameter
\begin{equation}
  \label{eq:Keldysh_parameter}
  \gamma = \frac{\omega_\mathrm{L} \sqrt{m E_\mathrm{g}}}{e F_0}.
\end{equation}
This parameter was introduced for a monochromatic laser field in the
approximation of parabolic bands.  In Eq.~\eqref{eq:Keldysh_parameter},
$\omega_\mathrm{L}$ is the laser frequency,
$m=(1/m_\mathrm{e}+1/m_\mathrm{h})^{-1}$ is the reduced mass of an
electron and a hole, $E_\mathrm{g} = \hbar \omega_\mathrm{g}$ is the
bandgap, $e$ is the absolute value of the electron charge, and $F_0$
is the amplitude of the laser field in the medium. In the limit $\gamma \gg 1$,
where the field is weak and/or its frequency is large, an electron is
excited from the valence band by absorbing an integer number of
photons in a way that is well described by conventional perturbation
theory. In the opposite extreme $\gamma \ll 1$, where the field is
strong and its frequency is sufficiently small, it is quantum
tunnelling that promotes electrons to the conduction band. Obviously,
this is the case if the external field is constant
($\omega_\mathrm{L}=0$), and interband tunnelling as a concept was
first introduced by Zener~\cite{Zener_PRSLA_1934} for time-independent
fields. In the tunnelling regime, the excitation rate is often assumed to
be determined by the instantaneous value of the electric field
$F_\mathrm{L}(t)$~\cite{Mitrofanov_PRL_2011}.

Electron dynamics in the intermediate regime $\gamma \sim 1$ are
particularly complex---a laser field cannot be treated as a small
perturbation in this regime, and, at the same time, interband
transitions cannot be described by pure tunnelling. A general
expression for the cycle-averaged excitation rate was given by
Keldysh in~\cite{Keldysh_JETP_1965}, but a completely
satisfactory description of electron dynamics within an optical cycle
of an ultrashort pulse is still missing. The dynamics can be modelled
by solving the time-dependent Schr\"odinger equation (TDSE) in a
suitable approximation, but there is no unambiguous way to distinguish
between valence- and conduction-band electrons as long as the external
field is present.  This is analogous to the problem of determining the
ionization rate of atoms and molecules in strong fields
\cite{Yakovlev_CP_2012}, for which many practical solutions exist, but
there is no mathematically rigorous solution in the case of a
few-cycle pulse. This motivated us to study the polarization response
of a dielectric in the regime $\gamma \sim 1$, as interband
transitions directly affect polarization and, at the same time,
polarization is a quantum-mechanical observable that determines all
optical properties of a solid.

\section{Theory}
We are mainly interested in the strong-field regime where
perturbation theory shows its shortcomings. In particular, the
polarization response of a solid to a laser pulse is no longer
described by a set of linear and nonlinear susceptibilities;
therefore, quantum-mechanical simulations are necessary to model
it. We solve the TDSE numerically in the velocity gauge, using the
method described in~\cite{Bachau_PRB_2006}. For each value of the
quasimomentum $\mathbf{k}$, the electron wave function is represented in
the basis of Bloch states $\ket{\phi_\mathbf{k}^n}$, which are evaluated
by solving the single-electron stationary Schr\"odinger equation with
an unperturbed Hamiltonian $\hat{H}_0$:
\begin{equation}
  \label{eq:stationary_SE}
  \hat{H}_0 \ket{\phi_\mathbf{k}^n} =
  E_\mathbf{k}^n \ket{\phi_\mathbf{k}^n}.
\end{equation}
Here, $n$ is a band index. This equation is solved in the basis of
$N_\mathrm{max}$ plane waves:
\begin{equation}
  \label{eq:plane_waves}
  \phi_\mathbf{k}^n(\mathbf{r}) = \sum_{j=1}^{N_\mathrm{max}} C_{\mathbf{k},j}^n
  \exp[\rmi (\mathbf{k}+\mathbf{K}_j) \mathbf{r}],
\end{equation}
where $\mathbf{K}_j$ denotes reciprocal lattice vectors, each of which
can be expressed as a combination of the primitive vectors
$\{\mathbf{b}_1,\mathbf{b}_2,\mathbf{b}_3\}$:
\begin{equation}
  \label{eq:K}
  \mathbf{K} = \sum_{\alpha=1}^3 m_\alpha \mathbf{b}_\alpha,\ m_\alpha \in \mathbb{Z}.
\end{equation}
Ansatz \eqref{eq:plane_waves} ensures that all Bloch states satisfy
$\phi_\mathbf{k}^n(\mathbf{r}+\mathbf{R}) =
\phi_\mathbf{k}^n(\mathbf{r}) \exp[\rmi \mathbf{k} \mathbf{R}]$,
where $\mathbf{R}$ is a vector in the direct Bravais lattice. This
property immediately follows from $\exp[\rmi \mathbf{K} \mathbf{R}] =
1$.

The coefficients in expansion \eqref{eq:plane_waves}, as well as
the energies $E_\mathbf{k}^n$, are determined by diagonalizing the
Hamiltonian $\hat{H}_0$ in the basis of plane waves $\exp[\rmi
\mathbf{K}_j \mathbf{r}]$. We assume that the interaction of a
particular electron with lattice ions and other electrons is
described by a pseudopotential $U(\mathbf{r})$ and expand this
pseudopotential in the plane-wave basis:
\begin{align}
  \label{eq:U_decomposition}
  U(\mathbf{r}) &= \sum_{j=1}^{N_\mathrm{max}} U_j \rme^{\rmi \mathbf{K}_j \mathbf{r}},\\
  U_{\mathbf{K}_j} &= \frac{1}{\Omega} \int_{\Omega} \rme^{-\rmi \mathbf{K}_j \mathbf{r}} U(\mathbf{r}) \,\rmd\mathbf{r},
\end{align}
where $\Omega$ is the volume of a unit cell. Writing the unperturbed Hamiltonian
as
\begin{equation}
  \label{eq:Hamiltionan}
  \hat{H}_0 = -\frac{\hbar^2}{2 m_0} \nabla^2 + U(\mathbf{r})
\end{equation}
and substituting ansatz \eqref{eq:plane_waves} into Eq.~\eqref{eq:stationary_SE},
we obtain the following eigensystem for the expansion coefficients
$C_{\mathbf{k},i}^n$ and energies $E_\mathbf{k}^n$:
\begin{equation}
  \label{eq:eigensystem}
  \sum_{j=1}^{N_\mathrm{max}} \left(
    \frac{\hbar^2}{2 m_0} (\mathbf{k}+\mathbf{K}_j)^2 \delta_{i j} +
    U_{\mathbf{K}_i-\mathbf{K}_j} \right) C_{\mathbf{k},j}^n =
  E_\mathbf{k}^n C_{\mathbf{k},i}^n.
\end{equation}
Here, $m_0$ is the electron rest mass, and the eigensystem can be
solved independently for each quasimomentum $\mathbf{k}$. The solutions
of this eigensystem are normalized to satisfy
\begin{equation}
  \label{eq:Bloch_normalization}
  \braket{\phi_\mathbf{k}^m}{\phi_\mathbf{k}^n} = 
  \frac{1}{\Omega} \int_{\Omega} \left[\phi_\mathbf{k}^m(\mathbf{r})\right]^*
  \phi_\mathbf{k}^n(\mathbf{r})\,\rmd^3\mathbf{r} =
  \delta_{m n},
\end{equation}
which implies that $\sum_j |C_{\mathbf{k},j}^n|^2 = 1$. Here and in the
following, a scalar product $\braket{\cdot}{\cdot}$ implies
integration over one unit cell.

Having evaluated Bloch states $\ket{\phi_\mathbf{k}^n}$, we use them
as a basis to solve the TDSE and thus model the interaction of
electrons with a laser pulse:
\begin{equation}
  \label{eq:TDSE}
  \rmi \hbar \pdiff{}{t} \ket{\psi_\mathbf{k}(t)} =
  \left(\hat{H}_0 + \hat{H}_\mathrm{int}(t)\right) \ket{\psi_\mathbf{k}(t)},
\end{equation}
where $\hat{H}_\mathrm{int}(t)$ is the interaction Hamiltonian. We
perform our simulations in the velocity gauge, where an external
electric field is described by the vector potential
$\mathbf{A}(\mathbf{r},t)$.  We also employ the dipole approximation:
$\mathbf{A}(\mathbf{r},t) \equiv \mathbf{A}(t)$. In SI units,
this corresponds to using the following form of the interaction
Hamiltonian:
\begin{equation}
  \label{eq:H_int}
  \hat{H}_\mathrm{int}(t) = \frac{e}{m_0} \mathbf{A}(t) \hat{\mathbf{p}}.
\end{equation}
Note that in CGS units the right hand side of Eq.~\eqref{eq:H_int}
must be divided by the vacuum speed of light $c$. The substitution of
the ansatz
\begin{equation}
  \label{eq:main_ansatz}
  \ket{\psi_\mathbf{k}(t)} =
  \sum_{n=1}^{N_\mathrm{max}} \alpha_\mathbf{k}^n(t) \ket{\phi_\mathbf{k}^n}
\end{equation}
into Eq.~\eqref{eq:TDSE} leads to the following system of
coupled differential equations:
\begin{equation}
  \label{eq:main_system}
  \rmi \hbar \pdiff{\alpha_\mathbf{k}^q(t)}{t} =
  E_\mathbf{k}^q \alpha_\mathbf{k}^q(t) +
  \frac{e}{m_0} \mathbf{A}(t) \sum_{l=1}^{N_\mathrm{max}}
  \mathbf{p}_\mathbf{k}^{q l} \alpha_\mathbf{k}^l(t).
\end{equation}
Here,
\begin{equation}
  \label{eq:momentum_ME}
  \mathbf{p}_\mathbf{k}^{q l} =
  \melement{\phi_\mathbf{k}^q}{\hat{\mathbf{p}}}{\phi_\mathbf{k}^l} =
  \hbar \sum_{j=1}^{N_\mathrm{max}} (\mathbf{k}+\mathbf{K}_j)
  \left(C_{\mathbf{k},j}^q\right)^* C_{\mathbf{k},j}^l 
\end{equation}
are the matrix elements of the momentum operator $\hat{\mathbf{p}}$. An important
advantage of the velocity gauge is that, as long as the dipole
approximation holds and electron scattering is neglected, there
is no coupling between different values of $\mathbf{k}$.
Mathematically, this is a consequence of
\begin{multline*}
  \int_{\mathbb{R}^3} \left(\phi_{\mathbf{k}'}^q(\mathbf{r})\right)^*
  \hat{\mathbf{p}}\, \phi_\mathbf{k}^l(\mathbf{r}) \,\rmd^3\mathbf{r} =\\=
  \melement{\phi_{\mathbf{k}'}^q}{\hat{\mathbf{p}}}{\phi_{\mathbf{k}}^l}
  \sum_{j} \rme^{\rmi (\mathbf{k}-\mathbf{k}') \mathbf{R}_j } \propto \delta_{\mathbf{k} \mathbf{k}'},
\end{multline*}
where the summation is performed over all the vectors of the Bravais
lattice $\mathbf{R}_j$, and we assume that both $\mathbf{k}$ and
$\mathbf{k}'$ belong to the first Brillouin zone. Due to this
property, Eqs.~\eqref{eq:main_system} can be solved independently
for each quasimomentum $\mathbf{k}$. According to our experience, the
downside of the velocity gauge is a larger number of bands required
for convergence, as compared to length-gauge simulations, where
different values of $\mathbf{k}$ are coupled to each other
\cite{Blount_1962,Glutsch_PRB_2004}, but even the two-band
approximation may be sufficient~\cite{Golde_PRB_2008}.

At the beginning of a simulation, all electrons are supposed to reside
in the valence bands (VB), while all the conduction bands (CB) are
empty. We solve Eqs.~\eqref{eq:main_system} for each of the
electrons independently and then add single-electron contributions
together to evaluate physical observables such as current density or
polarization. Obviously, our approach is only applicable if
(i) lattice vibrations have no significant effect on the investigated dynamics,
(ii) effects related to electron correlation and electron--hole
interaction are not important, and (iii) distortions of the
pseudopotential $U(\mathbf{r})$ upon electronic excitations are
negligible. While these approximations obviously have their limitations,
single-electron models proved to be useful for studying basic physics
related to phenomena occurring on attosecond and few-femtosecond time
scales~\cite{Bachau_PRB_2006,Kazansky_PRL_2009,Brabec_2008}.

The initial condition for solving Eqs.~\eqref{eq:main_system}
for a particular $\mathbf{k}$ is
\begin{equation}
  \label{eq:initial_condition}
  \alpha_\mathbf{k}^{q}(t_\mathrm{min}) = \delta_{q n}.
\end{equation}
Here, $n$ is a band where the electron was before the interaction
with a laser pulse.

In this paper, we are most interested in the \emph{macroscopic} electric current
density $\mathbf{j}(t)$ induced by the laser pulse
\begin{equation}
  \label{eq:j_definition}
  \mathbf{j}(t) = \sum_{n \in \mathrm{VB}} \int
  \mathbf{j}_{\mathbf{k},n}(t)\,\rmd^3\mathbf{k},
\end{equation}
where the integral is taken over the first Brillouin zone and
$\mathbf{j}_{\mathbf{k},n}(t)$ represents single-electron
contributions to the current density averaged over a unit cell:
\begin{multline}
  \label{eq:jk_definition}
  \mathbf{j}_{\mathbf{k},n}(t) = -\frac{e}{m_\mathrm{e}}
  \frac{1}{V} \int_{V} \rmd^3\mathbf{r}\,
  \biggl(
    \mbox{Re}\left[\psi_{\mathbf{k},n}^*(\mathbf{r},t)
    \hat{\mathbf{p}}
    \psi_{\mathbf{k},n}(\mathbf{r},t)\right] +\\+
    e \mathbf{A}(t) |\psi_{\mathbf{k},n}(\mathbf{r},t)|^2
  \biggr).
\end{multline}
Here and in the following, we add index $n$ to quantities related to
single-electron wave functions in order to indicate the initial
band. Thus, $\mathbf{j}_{\mathbf{k},n}(t)$ represents the contribution from
\emph{all} the bands to $\mathbf{j}(t)$ in a calculation where the electron
initially occupied band $n$. The term containing the vector potential
is required in the velocity gauge.

With the aid of Eq.~\eqref{eq:main_ansatz}, the single-electron current
density can be expressed via probability amplitudes
$\alpha_\mathbf{k}^q(t)$:
\begin{multline}
  \label{eq:jk_expanded}
  \mathbf{j}_{\mathbf{k},n}(t) = -\frac{e}{m_\mathrm{e}}
  \Biggl(e \mathbf{A}(t) +\\+ \mbox{Re}\biggl[
      \sum_{q,l} \left( \alpha_{\mathbf{k},n}^q(t) \right)^*
      \alpha_{\mathbf{k},n}^l(t) \mathbf{p}_\mathbf{k}^{q l} \biggr]
     \Biggr).
\end{multline}

Three different physical effects determine the induced currents:
polarization due to the displacement of valence-band electrons, the
light-driven motion of conduction-band electrons, and quantum beats
due to coherent superpositions of valence- and conduction-band
states. In the presence of a strong external electric field, these
contributions cannot be fully separated from each other, but the
overall induced current density $\mathbf{j}(t)$ is defined unambiguously. It is
this induced current that determines the change of the refractive
index and the generation of new frequency components. However, it is
conventional to describe optical response in terms of polarization,
which we define as
\begin{equation}
  \label{eq:polarization}
  \mathbf{P}(t) = \int_{-\infty}^t \mathbf{j}(t')\,\rmd t'.
\end{equation}
Note that the polarization defined this way is due to both bound and
free electrons. Ordinarily, $\mathbf{P}(t)$ would be assigned to bound
electrons, while $\mathbf{j}(t)$ would describe the motion of free
electrons. We make no attempt to make this distinction because, to the
best of our knowledge, there is no rigorous way to distinguish between
bound and free electrons as long as the external field is present
\cite{Yakovlev_CP_2012}. It is only after the laser pulse that
projecting a wave function onto conduction-band Bloch states yields
\emph{physically} relevant excitation probabilities.

For simplicity, we perform our numerical simulations in one spatial
dimension $x$, which we assume to be the polarization direction of the
laser field. Most of the equations in this section can be transformed
to their one-dimensional forms by replacing all vector quantities and
three-dimensional integrals with scalar quantities and one-dimensional
integrals: $\mathbf{r} \to x$, $\mathbf{k} \to k$. The only exception
is Eq.~\eqref{eq:j_definition}, where this procedure would result in
$j(t)$ being measured in wrong units. Reducing the dimensionality of
our original problem, we essentially assume that physical observables
do not depend on the coordinates $y$ and $z$, so that the integral over
the first Brillouin zone in Eq.~\eqref{eq:j_definition} reduces to
\begin{equation}
  \label{eq:j_1D}
  j(t) = \eta \sum_{n \in \mathrm{VB}} \int j_{k,n}(t)\,\rmd k,
\end{equation}
where the factor $\eta$, measured in $\mbox{m}^{-2}$, accounts for the
integration over $k_y$ and $k_z$. In practice, the value of $\eta$
should be chosen to approximate some known optical properties of
the solid, such as its refractive index or absorption.

From this point on, we will write our equations in atomic units
(at.\,u.), unless stated otherwise. In atomic units, $\hbar = e = m_0
= 1$, a unit of energy is equal to $27.21$~eV, and a unit of electric
field is equal to $5.142 \times 10^{11}\ \mbox{V}/\mbox{m}$. Because $\hbar=1$,
we will interchangeably use the words ``energy'' and ``frequency''.

\begin{figure}[htbp]
  \centering
  \includegraphics[width=0.45\textwidth]{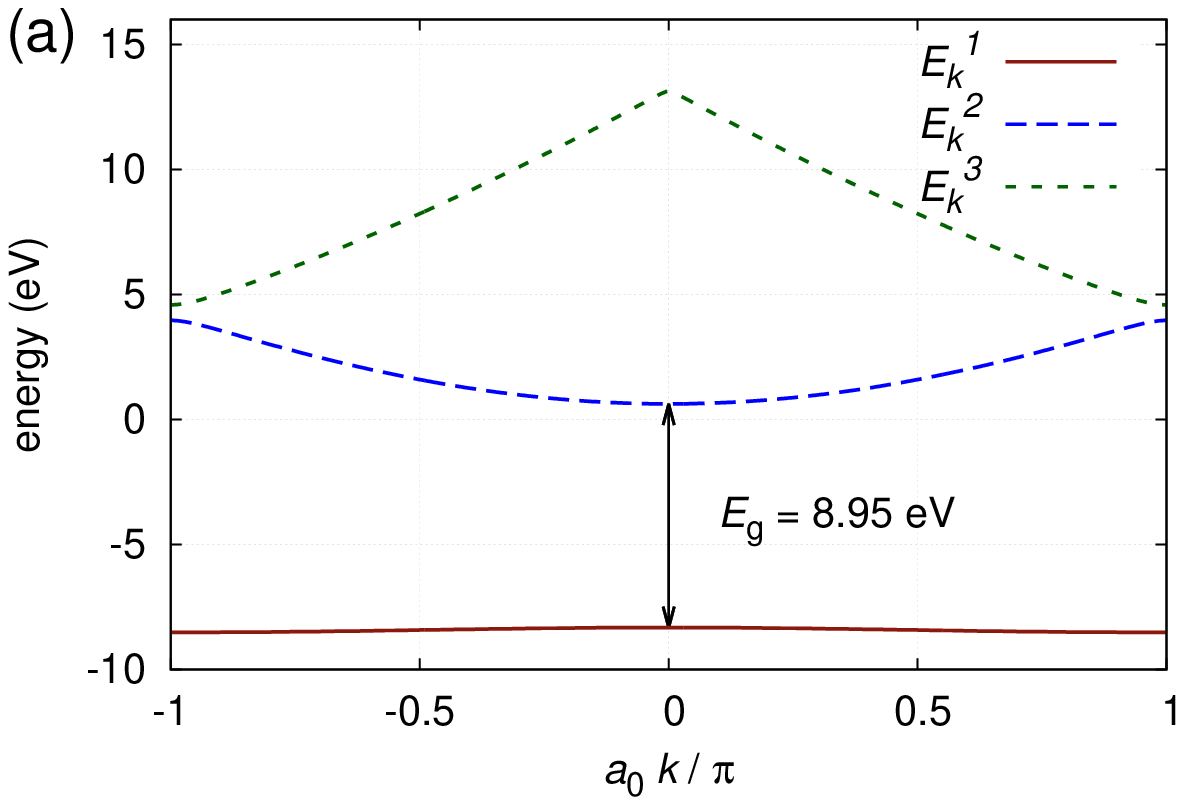}
  \includegraphics[width=0.45\textwidth]{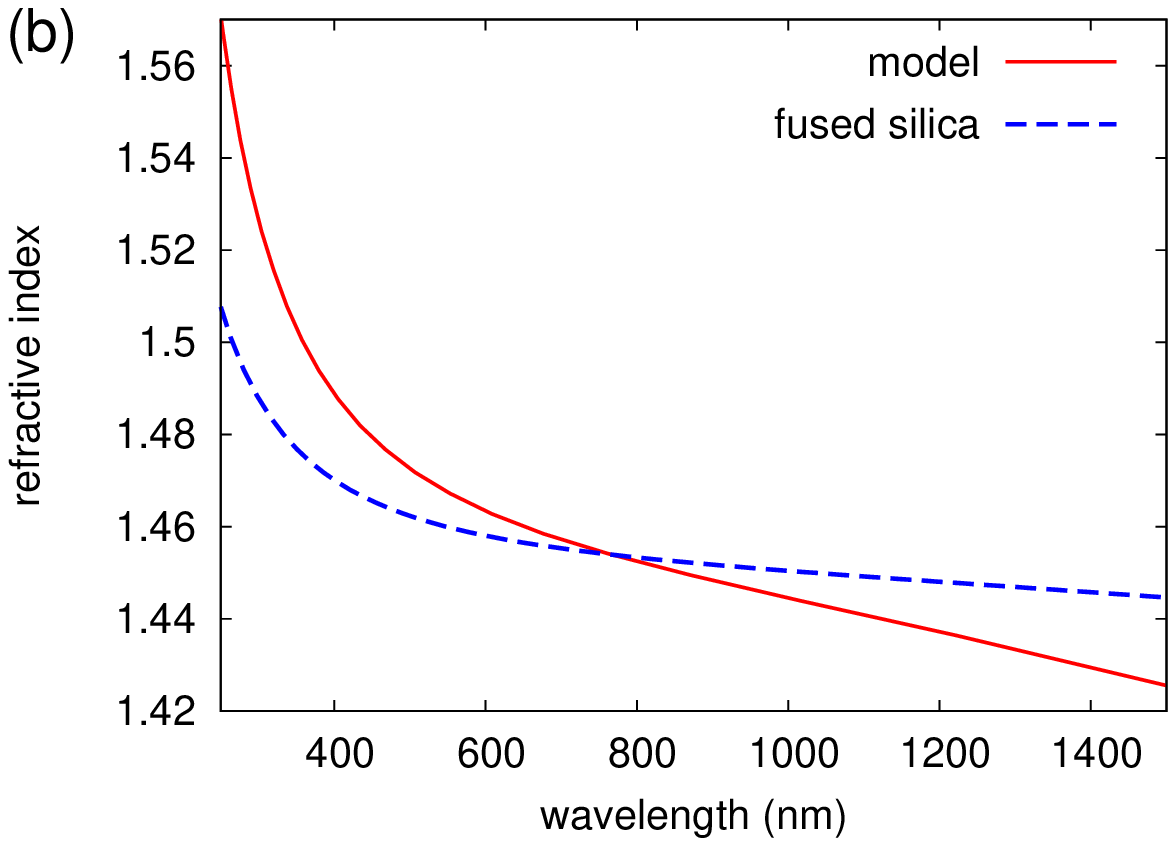}
  \caption{(a) The uppermost valence band and the two lowest conduction
    bands of the model potential \eqref{eq:potential}. (b) A comparison
    of the refractive index evaluated according to
    Eq.~\eqref{eq:refractive_index} with the refractive index of fused
    silica~\cite{Malitson_JOSA_1965}.}
  \label{fig:1}
\end{figure}

We adjusted our model potential $U(x)$ to obtain a bandgap close to
that of $\mbox{SiO}_2$: 8.95~eV \cite{Laughlin_PRB_1980}. To this end,
we set the lattice constant to $a_0 = 0.5\ \mbox{nm} = 9.45\
\mbox{at.}\,\mbox{u.}$, which is one of the lattice constants in
$\alpha$-quartz, and use the following expression for the central unit
cell:
\begin{multline}
  \label{eq:potential}
  U(x) = -0.7 \left(1 + \tanh[x+0.8]\right) \times\\ \times
  \left(1 + \tanh[-x+0.8]\right)
\end{multline}
for $|x| \le a_0/2$.
Outside the central unit cell, the potential is continued
periodically: $U(x+a_0)=U(x)$. The band structure that corresponds to
this potential is shown in Fig.~\ref{fig:1}(a).  At $k=0$ the energies
$E_k^n$ of the first four bands are $-35.48$, $-8.33$, $0.62$, and
$13.13$~eV. We regard the first two bands with negative values of
$E_k^n$ as valence bands (the lowest valence band is not shown
Fig.~\ref{fig:1}(a)), while all the bands with $E_k^n>0$ are regarded as
conduction bands.

We used the following expression for the vector potential of the
(linearly polarized) laser field acting on electrons in the solid:
\begin{equation}
  \label{eq:laser_pulse}
  A(t) = -\theta(\tau_\mathrm{L}-|t|)
  \frac{F_0}{\omega_\mathrm{L}}
  \cos^4\left(\frac{\pi t}{2 \tau_\mathrm{L}}\right)
  \sin(\omega_\mathrm{L} t + \varphi_\mathrm{CEP}).
\end{equation}
Here, $F_0$ is the amplitude of the electric field,
$\omega_\mathrm{L}$ is the central frequency of the laser pulse,
$\varphi_\mathrm{CEP}$ is its CEP,
$\theta(x)$ is the Heaviside step function, and $\tau_\mathrm{L}$ is
related to the full width at half maximum (FWHM) of $|A(t)|^2$ by
$\mbox{FWHM} = 4 \arccos\left(2^{-1/8}\right) \tau_\mathrm{L} / \pi
\approx 0.5224 \tau_\mathrm{L}$. The FWHM of $|A(t)|^2$ is very
close, although not precisely equal, to the FWHM of $|F(t)|^2$.
The external electric field acting on electrons is, by definition,
\begin{equation}
  \label{eq:electric_field}
  F(t) = -A'(t).
\end{equation}
Note that, in this work, we do not make any attempt to evaluate
the screening field created by the displacement of electrons---this field
is assumed to be a part of $A(t)$ and $F(t)$.

For a weak pulse, the polarization response is linear, so that, in the
frequency domain, $\tilde{P}(\omega) = \chi(\omega)
\tilde{F}(\omega)$, and the refractive index can be evaluated from the
linear susceptibility $\chi(\omega)$:
\begin{equation}
  \label{eq:refractive_index}
  n(\omega) = \sqrt{1 + \chi(\omega)}.
\end{equation}
In Fig.~\ref{fig:1}(b), we compare the refractive index of fused silica
with $n(\omega)$ evaluated using our model. The refractive index is
plotted against the laser wavelength $\lambda = 2 \pi c / \omega$. For
these simulations, we used several pulses with $F_0=10^{-5}\
\mbox{at.\,u.} = 5 \times 10^6\ \mbox{V}/\mbox{m}$,
$\mbox{FWHM}=4\pi/\omega_\mathrm{L}$, and values of
$\omega_\mathrm{L}$ that allowed us to cover the spectral range
presented in Fig.~\ref{fig:1}(b). Normalizing the polarization response,
we set $\eta = 0.111$ at.\,u.\ in Eq.~\eqref{eq:j_1D} and use this
value henceforth. Given the simplicity of our model, we find the
agreement with the measured refractive index very satisfactory.

The next section reports on simulations with much more intense fields,
where multiphoton excitations populate conduction bands. For those
simulations, we had to use 15 bands in order to achieve numerical
convergence in solving the TDSE.
A smaller number of bands results in discrepancies that are most
visible in the spectral range occupied by low-order harmonics, while
increasing the number of bands to 20 has a negligible effect on
the polarization response even for the most intense pulses
that we used in our modelling.

\section{Results and discussion}
\begin{figure*}[htbp]
  \centering
  \includegraphics[width=0.45\textwidth]{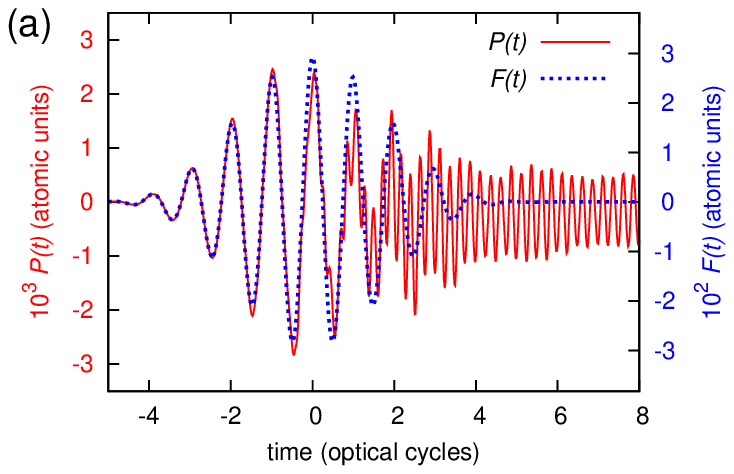}
  \includegraphics[width=0.45\textwidth]{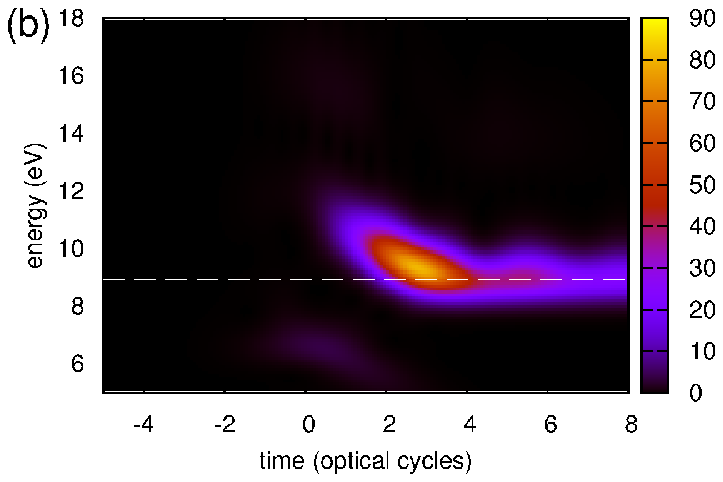}\\
  \includegraphics[width=0.45\textwidth]{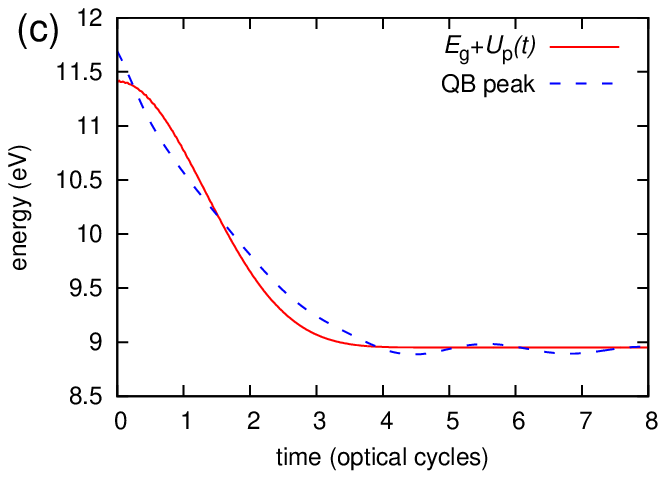}
  \includegraphics[width=0.45\textwidth]{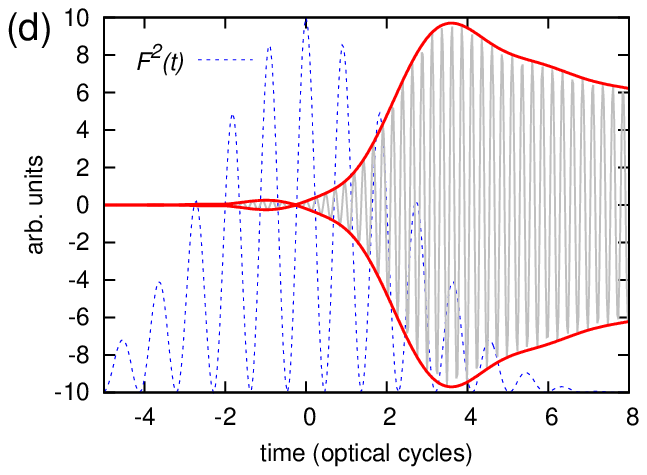}
  \caption{(a) The polarization $P(t)$ induced by a laser pulse with
    $\omega_\mathrm{L}=\omega_\mathrm{g}/4$ ($\lambda_\mathrm{L} =
    557\ \mbox{nm}$), a peak intensity of $3 \times 10^{13}\
    \mbox{W}/\mbox{cm}^2$, and $\varphi_\mathrm{CEP}=0$. The electric
    field $F(t)$ is shown by the dotted blue line; the FWHM of the
    pulse corresponds to the time range from $-1.5$ to $1.5$ optical
    cycles. (b) $S(t,\omega)$, which is the result of applying the
    wavelet transform to $P(t)$. The dashed white line marks the
    energy $E_\mathrm{g}=\hbar \omega_\mathrm{g}$. (c) The solid red
    line is the sum of the bandgap energy $E_\mathrm{g}$ and the
    ponderomotive energy $U_\mathrm{p}(t)$. For each time $t$, the dashed blue
    line depicts the above-bandgap frequency where $S(t,\omega)$ has a
    maximum.  (d) The quantum-beat part of the polarization response
    $\mathcal{F}^{-1}[\tilde{w}(\omega)\tilde{P}(\omega)]$
    (grey) and its envelope (red).
    The spectral window used for the analysis is given by Eq.~\eqref{eq:spectral_window}.
  }
  \label{fig:2}
\end{figure*}
For all the simulations in this section, we used pulses with the FWHM
equal to three optical oscillations: $\mbox{FWHM}=3 \times 2
\pi/\omega_\mathrm{L}$. The electric field $F(t)$ of such a pulse is
shown in Fig.~\ref{fig:2}(a) by the dotted blue line plotted against
the number of optical cycles $\omega_\mathrm{L} t/(2 \pi)$.  The solid
red line in this figure shows the polarization response $P(t)$
evaluated according to Eq.~\eqref{eq:polarization} for the case where the
central frequency is equal to a quarter of the bandgap
($\omega_\mathrm{L}=\omega_\mathrm{g}/4$, $\lambda_\mathrm{L}=2\pi c/
\omega_\mathrm{L} = 557$ nm), and the peak intensity is equal to
$I_\mathrm{L} = 3 \times 10^{13}\ \mbox{W}/\mbox{cm}^2$, which
corresponds to an amplitude of the electric field of $F_0 = 0.029\
\mbox{at.\,u.} = 1.5\ \mbox{V}/\mbox{\AA}$. The scales chosen for the
figure emphasize that, at the leading edge of the laser pulse, $P(t)$
is proportional to $F(t)$.  At later times, the polarization response
becomes increasingly nonlinear and high-frequency oscillations
appear, which persist even after the laser pulse ends. The frequency
of these oscillations is close to the bandgap frequency
$\omega_\mathrm{g} = E_\mathrm{g}/\hbar$. Thus, these are quantum
beats appearing due to the presence of coherent superpositions of
valence- and conduction-band states. This part of the polarization
response will be the main topic of our further discussion. In order to
distinguish it from conventional harmonics due to $\chi^{(3)}$,
$\chi^{(5)}$, $\chi^{(7)}, \ldots$ nonlinearities, we chose the
central laser frequency to be an even divisor of the band gap, taking
advantage of the fact that harmonics of even orders are absent if the
potential $U(x)$ possesses the inversion symmetry $U(-x)=U(x)$, like
our model potential \eqref{eq:potential}. Note that, due to the same
symmetry, the four-photon transition from the uppermost valence band
to the lowest conduction band at $k=0$ is forbidden, while interband
tunnelling can efficiently populate conduction-band states in the
middle of the Brillouin zone.
Indirectly, this is confirmed by the fact that the fast oscillations
in Fig.~\ref{fig:2}(a) only appear at very high intensities. At a peak
intensity of $I_\mathrm{L} = 10^{13}\ \mbox{W}/\mbox{cm}^2$, which is
just 3 times smaller than the one used for Fig.~\ref{fig:2}, the
amplitude of quantum-beat oscillations is smaller by approximately a
factor of 400, which corresponds to a drop of their intensity by five
orders of magnitude.

We investigate the quantum-beat part of the polarization response by
means of the wavelet analysis using the Morlet wavelet~\cite{Chui_1992}:
\begin{equation}
  \label{eq:wavelet}
  S(t,\omega) = \omega \left| \int_{-\infty}^\infty
    P(t') W\bigl(\omega (t' - t)\bigr)\,\rmd t' \right|^2,
\end{equation}
\begin{equation}
W(x) = \frac{1}{\sqrt{\tau_W}} \exp\left[\rmi x - \frac{x^2}{2 \tau_W^2}\right].
\end{equation}
The parameter $\tau_W$ determines the size of the temporal window, and
it must be chosen to provide an optimal compromise between the
temporal and spectral resolution. Our choice was $\tau_W =
3 \omega_\mathrm{g}/\omega_\mathrm{L}$.

The outcome of this analysis, $S(t,\omega)$, is shown as a
false-colour diagram in Fig.~\ref{fig:2}(b), the bandgap energy being
indicated by a dashed white line. For these parameters, the
quantum-beat signal completely dominates the third-order harmonic,
which is barely visible in the plot. In this time--frequency analysis,
the third-harmonic signal is centred at $6.7$~eV, and it is temporally
confined to the FWHM of the light pulse: $\omega_\mathrm{L} |t|/(2
\pi) \lesssim 1.5$.

One of the most striking features observed in Fig.~\ref{fig:2}(b) is
the fact that the quantum-beat signal initially appears at frequencies
exceeding $\omega_\mathrm{g}$ by roughly $2\ \mbox{eV}/\hbar$. The
instantaneous frequency of quantum beats then gradually decreases,
approaching $\omega_\mathrm{g}$ at the tailing edge of the laser
pulse.  In other words, the quantum-beat signal is negatively
chirped---its lower-frequency components are delayed with respect to
the higher-frequency ones. This cannot be explained by the presence of
the fifth-order harmonic---even though the fifth harmonic occupies the
relevant spectral range ($5 \hbar \omega_\mathrm{L} = 11.2\ \mbox{eV}$),
it must be confined to an even smaller temporal range
than the third harmonic, so that $S(t,\omega)$ can be considered free
from its contribution for $\omega_\mathrm{L} t/(2 \pi) \gtrsim 1$,
where most of the quantum-beat signal is observed.

\begin{figure*}[htbp]
  \centering
  \includegraphics[width=0.45\textwidth]{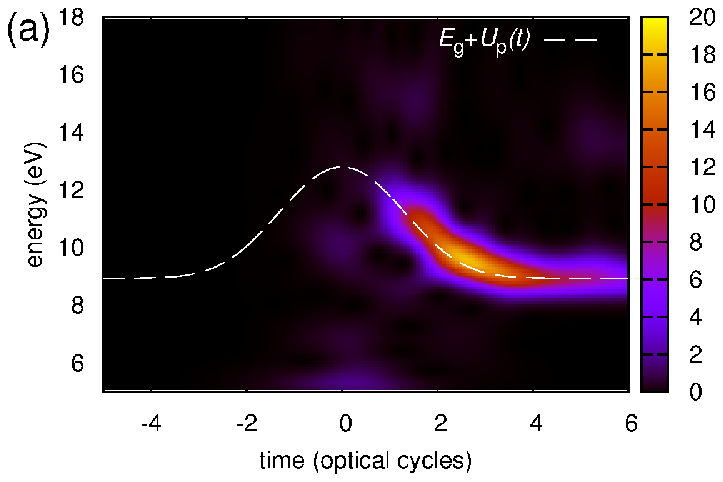}
  \includegraphics[width=0.45\textwidth]{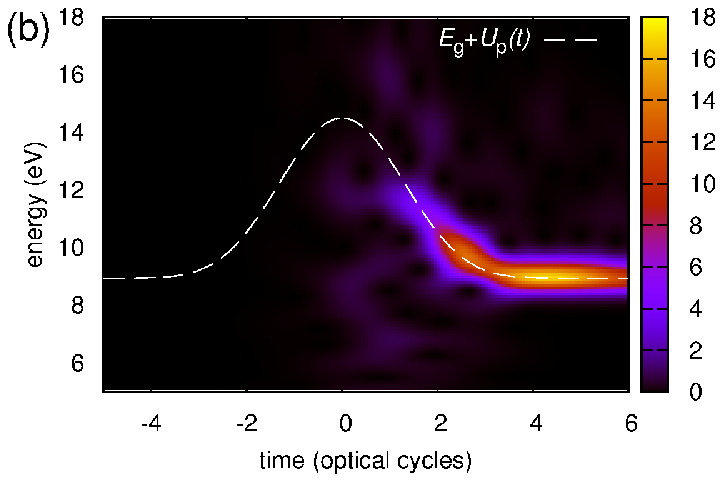}\\
  \includegraphics[width=0.45\textwidth]{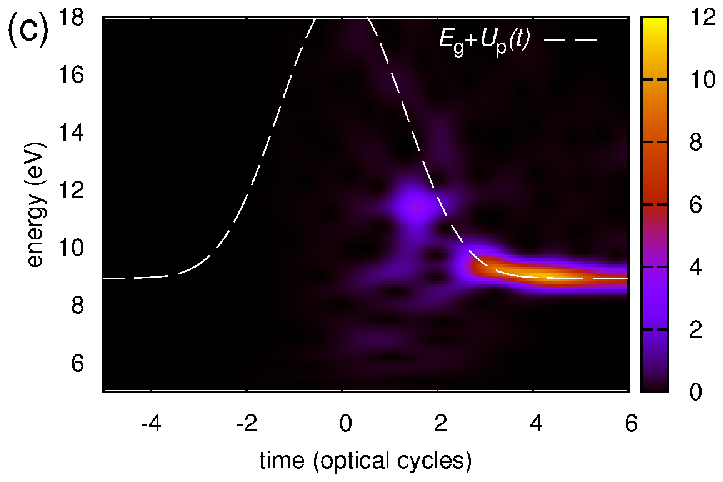}
  \includegraphics[width=0.45\textwidth]{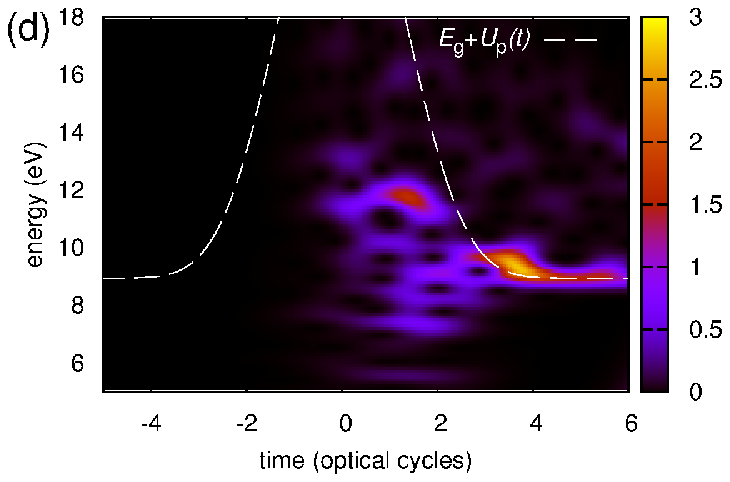}
  \caption{The wavelet analysis of the polarization induced by three-cycle
    pulses with a peak intensity of $3 \times 10^{13}\
    \mbox{W}/\mbox{cm}^2$, $\varphi_\mathrm{CEP}=0$, and the following
    central frequencies: (a) $\omega_\mathrm{L}=\omega_\mathrm{g}/5$
    ($\lambda_\mathrm{L} = 697\,\mbox{nm}$); (b)
    $\omega_\mathrm{L}=\omega_\mathrm{g}/6$ ($\lambda_\mathrm{L} =
    836\,\mbox{nm}$); (c) $\omega_\mathrm{L}=\omega_\mathrm{g}/8$
    ($\lambda_\mathrm{L} = 1108\,\mbox{nm}$); (d)
    $\omega_\mathrm{L}=\omega_\mathrm{g}/10$ ($\lambda_\mathrm{L} =
    1385\,\mbox{nm}$). The units in colour coding are arbitrary, but
    they are the same as those used for Fig.~\ref{fig:2}(b).}
  \label{fig:3}
\end{figure*}

We seek an explanation for the chirp of the quantum-beat signal in the
laser-driven motion of charge carriers. The average kinetic energy of
an electron--hole pair is the ponderomotive energy.  For a
monochromatic field with an amplitude $F_0$, the ponderomotive energy
is given, in the approximation of parabolic bands, by
\begin{equation}
  \label{eq:UP}
  U_\mathrm{p} = \frac{F_0^2}{4 m \omega_\mathrm{L}^2}.
\end{equation}
Here, $m$ is the reduced mass of an electron and a hole. We obtain the
effective masses of electrons and holes by fitting parabolas to the
lowest conduction and uppermost valence bands in the region
$|k|<\pi/(2 a_0)$, which yields $m_\mathrm{e}=0.34\ \mbox{at.\,u.}$
and $m_\mathrm{h}=3.61\ \mbox{at.\,u}$. These values are in good
agreement with those that can be found in the
literature~\cite{Schneider_PRL_1976,Chelikowsky_PRB_1977}.
Thus, the reduced effective
mass $m$ in Eq.~\eqref{eq:UP} is equal to $m=(1/m_\mathrm{e} +
1/m_\mathrm{h})^{-1}=0.35\ \mbox{at.\,u}$. The cycle-averaged total
energy of an electron--hole pair is equal to $E_\mathrm{g} + U_\mathrm{p}(t)$,
where $U_\mathrm{p}(t)$ is to be evaluated by replacing $F_0$ in Eq.~\eqref{eq:UP}
with the pulse envelope. This quantity is plotted as a red line
in Fig.~\ref{fig:2}(c). It is fairly close to the energies where
$S(t,\omega)$ has local maxima with respect to $\omega$. We observed
this kind of negative chirp in numerous simulations with different
parameters of the laser pulse, some of which are shown in
Fig.~\ref{fig:3}. Therefore, we conclude that the frequency of
quantum-beats exceeds $\omega_\mathrm{g}$ by approximately the
ponderomotive energy.

Fig.~\ref{fig:2}(d) presents a different kind of analysis that highlights
the quantum-beat part of the polarization response. In order to
suppress the third- and fifth-order harmonics, we multiplied the
Fourier transform of $P(t)$ with a soft spectral window that cuts all
frequency components that are further than $\omega_\mathrm{L}/2$ from
the bandgap frequency:
\begin{equation}
  \label{eq:spectral_window}
  \tilde{w}(\omega) = \theta\left(\frac{\omega_\mathrm{L}}{2} - \biggl||\omega|-\omega_\mathrm{g}\biggr|\right)
    \cos^2\left(\frac{\pi \biggl||\omega|-\omega_\mathrm{g}\biggr|}{\omega_\mathrm{L}}\right).
\end{equation}
The grey line in Fig.~\ref{fig:2}(d) shows the inverse Fourier transform
of $\tilde{w}(\omega) \tilde{P}(\omega)$. In other words, it
represents the part of the polarization response that oscillates at
frequencies close to the bandgap frequency. The envelope of these fast
oscillations is shown as a solid red line, and the dotted blue line
depicts the square of the laser field.  From this analysis, it is
obvious that the quantum-beat part of the polarization response
experiences its most rapid increase at a time that lies about two
optical cycles after the peak of the laser pulse. This is
counter-intuitive, as the quantum-beat signal is due to interband
excitations, and the rate of the interband excitations is expected to
reach its peak at the peak of the laser pulse. Another prominent
feature apparent from Fig.~\ref{fig:2}(d) is the decay of quantum beats
that begins at the trailing edge of the laser pulse. As our model does
not account for any relaxation phenomena (such as electron-phonon
interaction or spontaneous emission of radiation), the origin of this
decay lies in the dephasing of dipole oscillators associated with
each pair of coherently populated valence- and conduction-band
states. Each such coherent superposition creates a current $j_{k}(t)$
described by Eq.~\eqref{eq:jk_expanded}. If the laser pulse ends at a
time $t_1$, then $\alpha_k^q(t) = \alpha_k^q(t_1) \exp[-\rmi (t-t_1)
E_k^q]$, so that
\begin{equation}
  \label{eq:jk_field_free}
  j_k(t) = -\mbox{Re}
    \sum_{q,l} \left( \alpha_k^q(t_1) \right)^*
    \alpha_k^l(t_1) p_k^{q l} \rme^{\rmi (t-t_1) (E_k^q-E_k^l)} .
\end{equation}
For brevity, we have omitted the index specifying the initial band of
the electron. For a given pair of bands $q \ne l$,
Eq.~\eqref{eq:jk_field_free} describes a current oscillating with a
$k$-dependent frequency $E_k^q-E_k^l$. Because of this dependence, the
net current density \eqref{eq:j_1D} attenuates as most currents
$j_k(t)$ get out of phase with each other.

Fig.~\ref{fig:3} presents the wavelet analysis of polarization
response functions for several other values of the central laser frequency
$\omega_\mathrm{L}$. The dashed white lines depict $E_\mathrm{g}
+ U_\mathrm{p}(t)$. In Figs.~\ref{fig:3}(a) and (b), these lines closely
follow local maxima of $S(t,\omega)$. For smaller laser
frequencies, the ponderomotive potential $U_\mathrm{p}$ significantly exceeds
the width of the lowest conduction band (see Fig.~\ref{fig:1}(a)). This
is the regime where Bloch oscillations~\cite{Bloch_ZPA_1929} must play an
important role~\cite{Ghimire_Nature-Physics_2011}.

In Fig.~\ref{fig:3}(a), where valence-band electrons can be excited by
five-photon absorption, we made an exception from our choice to use
those values of
$\omega_\mathrm{L}$ that are even divisors of $\omega_\mathrm{g}$. Still, the
time--frequency analysis looks very similar to that for
$\omega_\mathrm{L}=\omega_\mathrm{g}/4$, shown in Fig.~\ref{fig:2}(b), and the
contribution from the fifth harmonic generation is so weak that it is
practically invisible in our colour scheme. For all the cases shown in
Fig.~\ref{fig:3}, we observe the same qualitative features as those in
Fig.~\ref{fig:2}(b): the frequency components of the polarization response that lie
above the bandgap appear delayed with respect to the peak of the laser
pulse, and the instantaneous frequency of these oscillations decreases
with time, eventually approaching $\omega_\mathrm{g}$. We also observe a rapid
decrease in the magnitude of $S(t,\omega)$.

\begin{figure*}[htbp]
  \centering
  \includegraphics[width=0.45\textwidth]{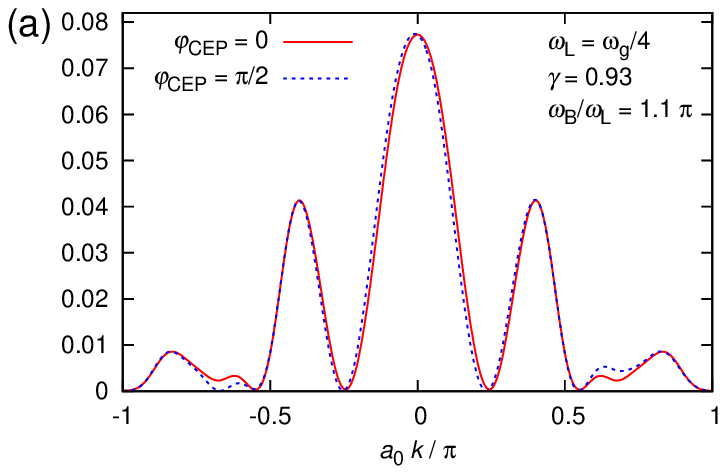}
  \includegraphics[width=0.45\textwidth]{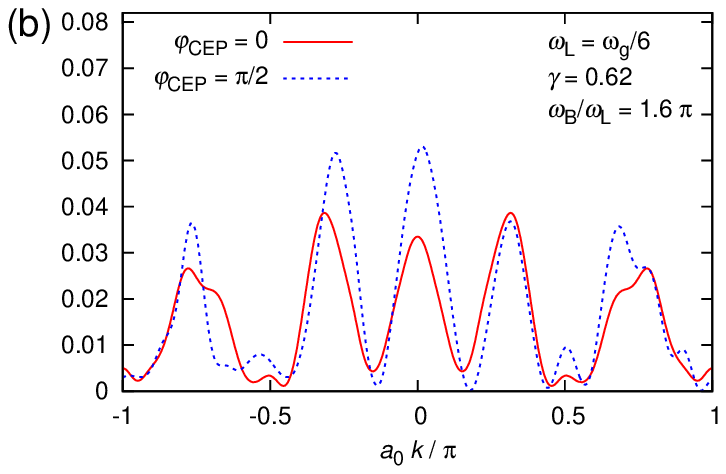}\\
  \includegraphics[width=0.45\textwidth]{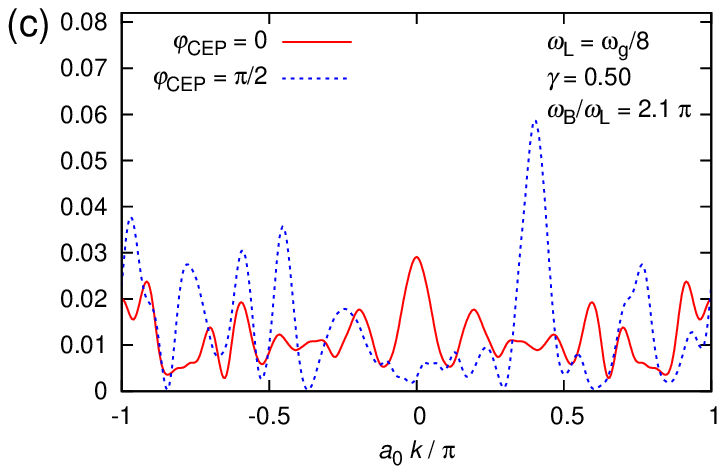}
  \includegraphics[width=0.45\textwidth]{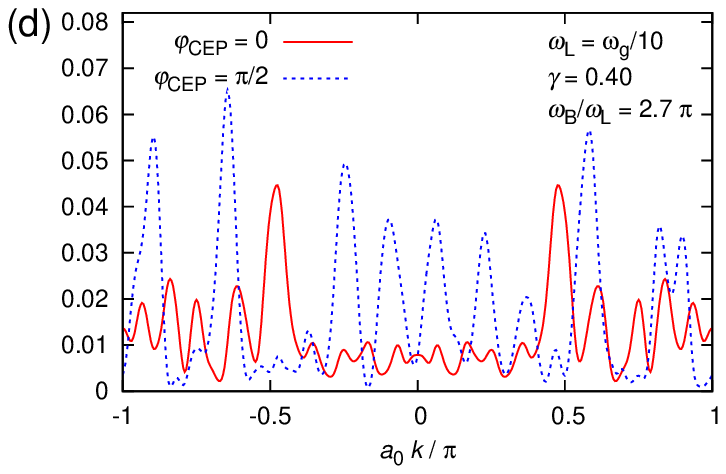}
  \caption{The probability for an electron that initially occupies a
    state in the uppermost valence band with a certain quasimomentum
    $k$ to be excited to one of the conduction bands by the end of a
    three-cycle laser pulse with a peak intensity of $3 \times 10^{13}\
    \mbox{W}/\mbox{cm}^2$. Results obtained for cosine pulses
    ($\varphi_\mathrm{CEP}=0$, solid red lines) are compared with
    those for sine pulses ($\varphi_\mathrm{CEP}=\pi/2$, dotted blue
    lines). The central laser frequency $\omega_\mathrm{L}$, the
    Keldysh parameter $\gamma$, and the ratio of the Bloch frequency
    $\omega_\mathrm{B}$ to the laser frequency are specified in the
    plots.}
  \label{fig:4}
\end{figure*}

To further understand the origin of these trends, we plot the sum of
all conduction-band populations at the end of our simulations in
Fig.~\ref{fig:4}. These are the probabilities to find an electron with
a given quasimomentum $k$ in one of the conduction bands at a time $t
> \tau_\mathrm{L}$ where the laser field is absent (since we do not
account for relaxation phenomena, populations do not change in the
absence of external fields). The distribution of excited electrons
dramatically changes as $\omega_\mathrm{L}$ decreases from $\omega_\mathrm{g}/4$
to $\omega_\mathrm{g}/10$: the pronounced peak at $k=0$ disappears, and many
irregular peaks appear across the Brillouin zone. At the same time,
the distributions become increasingly sensitive to the
CEP of the laser pulse. We consider two physical
phenomena that may be responsible for this trend: interband tunnelling
and Bloch oscillations.

The first phenomenon, interband tunnelling, represents an extreme
regime of interband excitations where the Keldysh parameter $\gamma$
is much smaller than 1. In this regime, interband transitions should
be viewed not as a result of absorbing a certain number of photons,
but as a result of quantum tunnelling~\cite{Zener_PRSLA_1934}. Also,
in the tunnelling regime, the CEP is an important
parameter because it is the field of the laser pulse, rather than its
envelope, that controls the tunnelling rate. According to
Fig.~\ref{fig:4} and all the other similar simulations that we have
performed, the onset of the CEP dependence coincides with $\gamma$
decreasing below a value approximately equal to $0.7$. Furthermore,
the tunnelling picture provides an intuitive explanation for the
multitude of peaks in Figs.~\ref{fig:4}(c) and (d): they may be
attributed to the interference among electron wave packets launched to
conduction bands within different half-cycles of the laser pulse,
which is analogous to the above-threshold ionization spectra of atoms
exposed to intense laser pulses~\cite{Delone_2000}.

What the tunnelling picture alone cannot explain is the disappearance
of the peak at $k=0$. Indeed, all analytical models for the tunnelling
rate predict that it should rapidly decrease as the bandgap increases
\cite{Zener_PRSLA_1934,Keldysh_JETP_1965,Kane_JPCS_1960}, and our
model potential has the smallest bandgap at $k=0$. Therefore, the
motion of excited electrons in conduction bands must play a
significant role~\cite{Golde_PRB_2008}, Bloch oscillations being the
most prominent manifestation of this motion.  An electron excited to
the bottom of a band at a zero-crossing of the vector
potential ($A(t_0)=0$) may reach the boundary of the first Brillouin
zone if, at some later moment $t_1$, the condition $|A(t_1)|=\pi/a_0$
is satisfied. Thus, Bloch oscillations should be considered important
if the amplitude of the vector potential $F_0/\omega_\mathrm{L}$ is
comparable to $\pi/a_0$.  Introducing the Bloch frequency
\begin{equation}
  \label{eq:Bloch_frequency}
  \omega_\mathrm{B} = \frac{e a_0 F_0}{\hbar}\ \mbox{(SI units)},
\end{equation}
this condition can be expressed as $\omega_\mathrm{B}/\omega_\mathrm{L} = \pi$.
The labels in Figs.~\ref{fig:4}(a-d) provide values of
$\omega_\mathrm{B}/\omega_\mathrm{L}$ in units of $\pi$ for all the
four simulations.  The appearance of the CEP dependence and the complex
structures in the distribution of conduction-band electrons coincide
not only with $\gamma$ becoming smaller than $1$, but also with
$\omega_\mathrm{B}/\omega_\mathrm{L}$ becoming larger than $\pi$. In
fact, this is not a coincidence, as
\begin{equation}
  \frac{\omega_\mathrm{B}}{\omega_\mathrm{L}} =
  \frac{a_0 \sqrt{m E_\mathrm{g}}}{\hbar \gamma}\ \mbox{(SI units)},
\end{equation}
as follows from Eqs.~\eqref{eq:Keldysh_parameter} and
\eqref{eq:Bloch_frequency}. For our parameters,
$\omega_\mathrm{B}/\omega_\mathrm{L} \approx 1.02 \pi / \gamma$, so
that the onset of tunnelling coincides with the onset of
Bloch oscillations, irrespectively of the laser pulse parameters.

\begin{figure*}[htbp]
  \centering
  \includegraphics[width=0.45\textwidth]{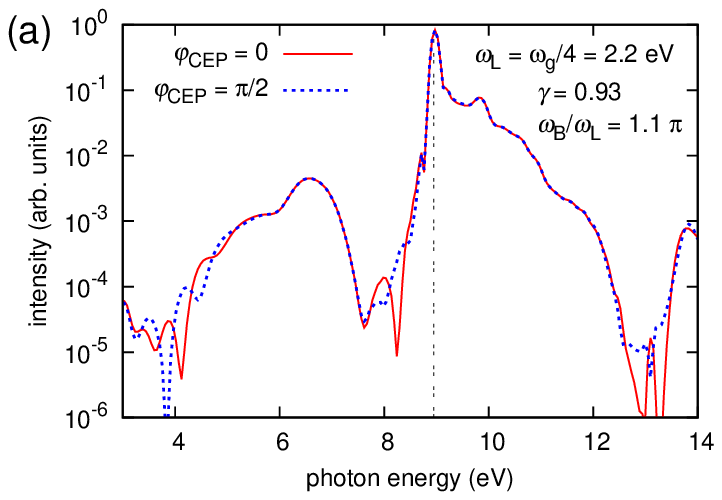}
  \includegraphics[width=0.45\textwidth]{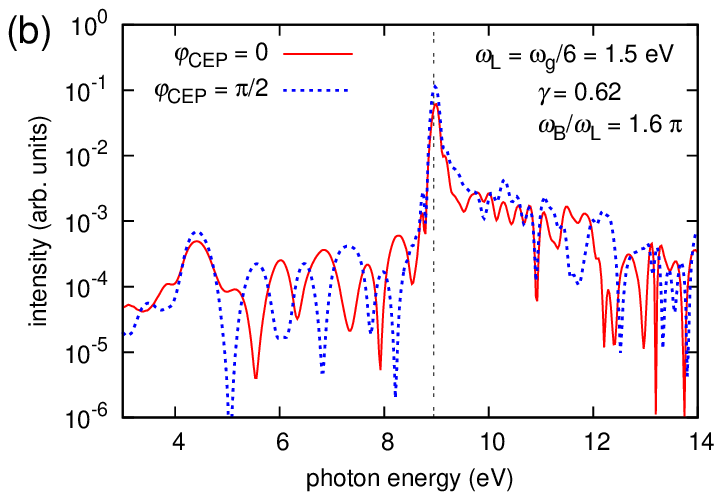}\\
  \includegraphics[width=0.45\textwidth]{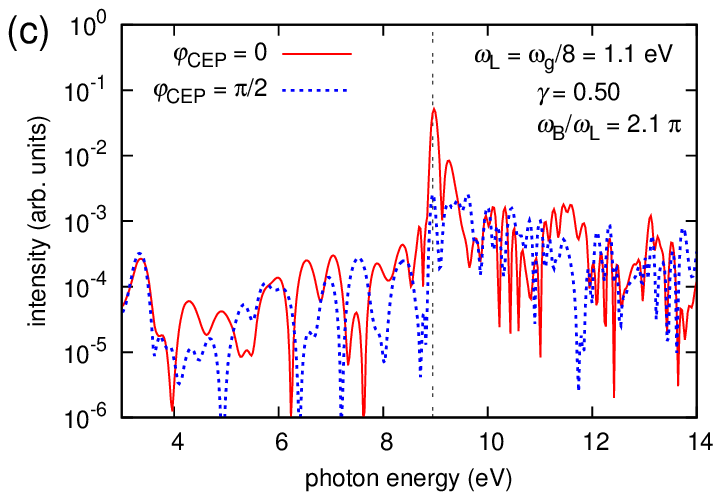}
  \includegraphics[width=0.45\textwidth]{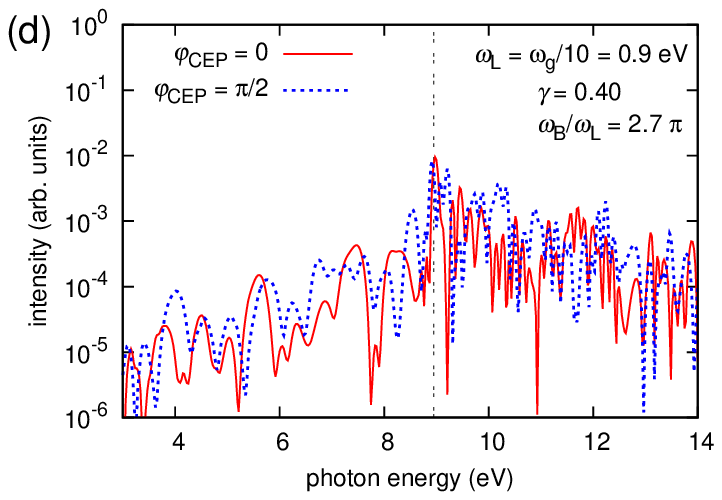}
  \caption{The power spectra of the polarization response for
    $\varphi_\mathrm{CEP}=0$ (solid red lines) and
    $\varphi_\mathrm{CEP}=\pi/2$ (dotted blue lines). The parameters
    of the laser pulse are the same as those used for
    Figs.~\ref{fig:4}(a-d), the values of the laser frequency
    $\omega_\mathrm{L}$ are specified in plot labels. The
    double-dashed black vertical lines mark the bandgap energy.
    Note that a part of the quantum-beat spectrum appears below the
    bandgap.}
  \label{fig:5}
\end{figure*}

Without interband excitations, Bloch oscillations obviously cannot
occur, but is it the nature of interband transitions or the Bloch
oscillations themselves that are mainly responsible for the observed
CEP dependencies? While we cannot answer this question with certainty,
we can argue that interband tunnelling alone is not sufficient for a
polarization response to be CEP dependent. Indeed, tunnelling must be
the dominant excitation mechanism in the parameter regime of
Fig.~\ref{fig:4}(a) because, in spite of the forbidden four-photon
transitions at $k=0$, the bottom of the conduction band is strongly
populated after the laser pulse. However, the polarization response is almost
CEP independent. Therefore, our simulations with
$\omega_\mathrm{L}=\omega_\mathrm{g}/4$ provide an example where the dominance of
interband tunnelling over (perturbative) multiphoton excitations does not lead to a
significant CEP dependence.

The quantum-mechanical observables that we have discussed so far
cannot be directly measured in a realistic experiment. The
most accessible quantity for experiments is the spectrum of emitted
radiation. In order to evaluate such spectra, one needs to account for
absorption and phase-matching~\cite{Ghimire_PRA_2012}, which is beyond
of scope of this work. Nevertheless, important conclusions about these
spectra can be made by analyzing the Fourier transform of the
polarization response $P(t)$ that we evaluate with our model. In
Fig.~\ref{fig:5}, we plot polarization spectra for the same
simulations as those presented in Fig.~\ref{fig:4}. We see that the
spectra are also sensitive to the CEP of the laser
pulse. Similarly to excitation probabilities, the CEP dependence is
very weak for $\omega_\mathrm{L}=\omega_\mathrm{g}/4$, but it becomes very
pronounced already at $\omega_\mathrm{L}=\omega_\mathrm{g}/6$, and it affects not
only the quantum-beat part of the polarization response, but also the
conventional harmonics of third, fifth etc orders.

Consistent with the results shown in Fig.~\ref{fig:3}, the
quantum-beat signal rapidly decreases with decreasing
$\omega_\mathrm{L}$. At the same time, according to Fig.~\ref{fig:4},
the excitation probabilities are comparable in all the four
cases. Consequently, the decrease in the polarization response is due
to dephasing. Most importantly, we used longer pulses for longer
wavelengths in order to have the same number of optical cycles in all
the pulses. As a result, dephasing played a more significant role for
the 14-fs pulse with $\omega_\mathrm{L}=\omega_\mathrm{g}/10$ than for the 5.5-fs
pulse with $\omega_\mathrm{L}=\omega_\mathrm{g}/4$. In addition to that, the
dephasing time is inversely proportional to the spectral width of the
excitation, and excited electrons occupy a larger part of the
Brillouin zone in the tunnelling regime (see Fig.~\ref{fig:4}).


\section{Conclusions and outlook}
We have analyzed the polarization response of a model dielectric
resembling SiO${}_2$ to few-cycle laser pulses that are strong enough
to efficiently excite electrons from valence to conduction
bands. These interband transitions create coherent superpositions of
states that manifest themselves as quantum beats in the polarization.
Obviously, the dynamics of interband excitations and dephasing
determine the properties of the quantum-beat part of the polarization
response. It is less obvious that the instantaneous frequency of
quantum-beat oscillations changes with time, approaching the bandgap
frequency as the laser field attenuates. We find that, during the
laser pulse, the quantum-beat frequency (averaged over a laser cycle)
is larger than the bandgap frequency by $U_\mathrm{p}/\hbar$, unless the
ponderomotive energy $U_\mathrm{p}$ exceeds the width of the lowest conduction
band. Thus, the emitted burst of high-frequency radiation is
negatively chirped. We have also observed that quantum beats appear in
the polarization response significantly delayed with respect to the
peak of a laser pulse, even though most electron--hole pairs must
appear when the laser field is strongest. Apparently, the effect of a
strong field on charge carriers suppresses
single-photon transitions from conduction- to valence-band states.
A satisfactory explanation of this
phenomenon probably requires the development of an analytical theory,
which may be triggered by this work.

For a dielectric like SiO${}_2$, the onset of interband tunnelling
($\gamma \sim 1$) coincides with the onset of Bloch oscillations
($\omega_\mathrm{B} \sim \pi \omega_\mathrm{L}$). In the tunnelling
regime, we have observed that the nonlinear polarization and the final
distribution of charge carriers are sensitive to the CEP of a laser pulse.
In particular, the spectrum of the emitted
high-frequency radiation depends on the CEP, which should be possible
to observe in experiments. Such a measurement would have to overcome
obstacles related to the strong absorption above the bandgap and the
phase mismatch below the bandgap~\cite{Ghimire_PRA_2012}. Also, it
must be mentioned that the coherent few-femtosecond pulse radiated by
quantum beats spectrally overlaps with the incoherent radiation due to
fluorescence, which our model does not account for. Some ultrashort temporal
gating may therefore be necessary to isolate the radiation due to quantum
beats. Still, in spite of all these obstacles, such measurements
should be feasible at the current stage of technology.

\section*{Acknowledgements}
The authors gratefully acknowledge fruitful discussions with
Prof.\ F.~Krausz and Prof.\ K.~Yabana. This work was supported by the DFG Cluster
of Excellence: Munich-Centre for Advanced Photonics.

\bibliography{references}

\begin{thebibliography}{10}%
\makeatletter
\providecommand \@ifxundefined [1]{%
 \ifx #1\undefined \expandafter \@firstoftwo
 \else \expandafter \@secondoftwo
\fi
}%
\providecommand \@ifnum [1]{%
 \ifnum #1\expandafter \@firstoftwo
 \else \expandafter \@secondoftwo
\fi
}%
\providecommand \enquote [1]{``#1''}%
\providecommand \bibnamefont  [1]{#1}%
\providecommand \bibfnamefont [1]{#1}%
\providecommand \citenamefont [1]{#1}%
\providecommand\href[0]{\@sanitize\@href}%
\providecommand\@href[1]{\endgroup\@@startlink{#1}\endgroup\@@href}%
\providecommand\@@href[1]{#1\@@endlink}%
\providecommand \@sanitize [0]{\begingroup\catcode`\&12\catcode`\#12\relax}%
\@ifxundefined \pdfoutput {\@firstoftwo}{%
 \@ifnum{\z@=\pdfoutput}{\@firstoftwo}{\@secondoftwo}%
}{%
 \providecommand\@@startlink[1]{\leavevmode\special{html:<a href="#1">}}%
 \providecommand\@@endlink[0]{\special{html:</a>}}%
}{%
 \providecommand\@@startlink[1]{%
  \leavevmode
  \pdfstartlink
   attr{/Border[0 0 1 ]/H/I/C[0 1 1]}%
   user{/Subtype/Link/A<</Type/Action/S/URI/URI(#1)>>}%
  \relax
 }%
 \providecommand\@@endlink[0]{\pdfendlink}%
}%
\providecommand \url  [0]{\begingroup\@sanitize \@url }%
\providecommand \@url [1]{\endgroup\@href {#1}{\urlprefix}}%
\providecommand \urlprefix [0]{URL }%
\providecommand \Eprint[0]{\href }%
\@ifxundefined \urlstyle {%
  \providecommand \doi [1]{doi:\discretionary{}{}{}#1}%
}{%
  \providecommand \doi [0]{doi:\discretionary{}{}{}\begingroup
  \urlstyle{rm}\Url }%
}%
\providecommand \doibase [0]{http://dx.doi.org/}%
\providecommand \Doi[1]{\href{\doibase#1}}%
\providecommand \bibAnnote [3]{%
  \BibitemShut{#1}%
  \begin{quotation}\noindent
    \textsc{Key:}\ #2\\\textsc{Annotation:}\ #3%
  \end{quotation}%
}%
\providecommand \bibAnnoteFile [2]{%
  \IfFileExists{#2}{\bibAnnote {#1} {#2} {\input{#2}}}{}%
}%
\providecommand \typeout [0]{\immediate \write \m@ne }%
\providecommand \selectlanguage [0]{\@gobble}%
\providecommand \bibinfo [0]{\@secondoftwo}%
\providecommand \bibfield [0]{\@secondoftwo}%
\providecommand \translation [1]{[#1]}%
\providecommand \BibitemOpen[0]{}%
\providecommand \bibitemStop [0]{}%
\providecommand \bibitemNoStop [0]{.\EOS\space}%
\providecommand \EOS [0]{\spacefactor3000\relax}%
\providecommand \BibitemShut [1]{\csname bibitem#1\endcsname}%
\bibitem{Stuart_PRB_1996}%
  \BibitemOpen
  \bibfield{author}{%
  \bibinfo {author} {\bibfnamefont{B.~C.}\ \bibnamefont{Stuart}}, \bibinfo
  {author} {\bibfnamefont{M.~D.}\ \bibnamefont{Feit}}, \bibinfo {author}
  {\bibfnamefont{S.}~\bibnamefont{Herman}}, \bibinfo {author}
  {\bibfnamefont{A.~M.}\ \bibnamefont{Rubenchik}}, \bibinfo {author}
  {\bibfnamefont{B.~W.}\ \bibnamefont{Shore}},\ and\ \bibinfo {author}
  {\bibfnamefont{M.~D.}\ \bibnamefont{Perry}},\ }%
  \bibfield{journal}{%
  \Doi{10.1103/PhysRevB.53.1749}{\bibinfo {journal} {Phys. Rev. B}}\ }%
  \textbf{\bibinfo {volume} {53}},\ \bibinfo {pages} {1749} (\bibinfo {year}
  {1996})%
  \bibAnnoteFile{NoStop}{Stuart_PRB_1996}%
\bibitem{Lenzner_PRL_1998}%
  \BibitemOpen
  \bibfield{author}{%
  \bibinfo {author} {\bibfnamefont{M.}~\bibnamefont{Lenzner}}, \bibinfo
  {author} {\bibfnamefont{J.}~\bibnamefont{Kr\"uger}}, \bibinfo {author}
  {\bibfnamefont{S.}~\bibnamefont{Sartania}}, \bibinfo {author}
  {\bibfnamefont{Z.}~\bibnamefont{Cheng}}, \bibinfo {author}
  {\bibfnamefont{C.}~\bibnamefont{Spielmann}}, \bibinfo {author}
  {\bibfnamefont{G.}~\bibnamefont{Mourou}}, \bibinfo {author}
  {\bibfnamefont{W.}~\bibnamefont{Kautek}},\ and\ \bibinfo {author}
  {\bibfnamefont{F.}~\bibnamefont{Krausz}},\ }%
  \bibfield{journal}{%
  \Doi{10.1103/PhysRevLett.80.4076}{\bibinfo {journal} {Phys. Rev. Lett.}}\ }%
  \textbf{\bibinfo {volume} {80}},\ \bibinfo {pages} {4076} (\bibinfo {year}
  {1998})%
  \bibAnnoteFile{NoStop}{Lenzner_PRL_1998}%
\bibitem{Mao_APA_2004}%
  \BibitemOpen
  \bibfield{author}{%
  \bibinfo {author} {\bibfnamefont{S.~S.}\ \bibnamefont{Mao}}, \bibinfo
  {author} {\bibfnamefont{F.}~\bibnamefont{Qu\'{e}r\'{e}}}, \bibinfo {author}
  {\bibfnamefont{S.}~\bibnamefont{Guizard}}, \bibinfo {author}
  {\bibfnamefont{X.}~\bibnamefont{Mao}}, \bibinfo {author}
  {\bibfnamefont{R.~E.}\ \bibnamefont{Russo}}, \bibinfo {author}
  {\bibfnamefont{G.}~\bibnamefont{Petite}},\ and\ \bibinfo {author}
  {\bibfnamefont{P.}~\bibnamefont{Martin}},\ }%
  \bibfield{journal}{%
  \Doi{10.1007/s00339-004-2684-0}{\bibinfo {journal} {Applied Physics A}}\ }%
  \textbf{\bibinfo {volume} {79}},\ \bibinfo {pages} {1695} (\bibinfo {year}
  {2004})%
  \bibAnnoteFile{NoStop}{Mao_APA_2004}%
\bibitem{Hentschel_Nature_2001}%
  \BibitemOpen
  \bibfield{author}{%
  \bibinfo {author} {\bibfnamefont{M.}~\bibnamefont{Hentschel}}, \bibinfo
  {author} {\bibfnamefont{R.}~\bibnamefont{Kienberger}}, \bibinfo {author}
  {\bibfnamefont{C.}~\bibnamefont{Spielmann}}, \bibinfo {author}
  {\bibfnamefont{G.}~\bibnamefont{Reider}},\ and\ \bibinfo {author}
  {\bibfnamefont{N.}~\bibnamefont{Milosevic}},\ }%
  \bibfield{journal}{%
  \Doi{10.1038/35107000}{\bibinfo {journal} {Nature}}\ }%
  \textbf{\bibinfo {volume} {414}},\ \bibinfo {pages} {509} (\bibinfo {year}
  {2001})%
  \bibAnnoteFile{NoStop}{Hentschel_Nature_2001}%
\bibitem{Krausz_RMP_2009}%
  \BibitemOpen
  \bibfield{author}{%
  \bibinfo {author} {\bibfnamefont{F.}~\bibnamefont{Krausz}}\ and\ \bibinfo
  {author} {\bibfnamefont{M.}~\bibnamefont{Ivanov}},\ }%
  \bibfield{journal}{%
  \Doi{10.1103/RevModPhys.81.163}{\bibinfo {journal} {Rev. Mod. Phys.}}\ }%
  \textbf{\bibinfo {volume} {81}},\ \bibinfo {pages} {163} (\bibinfo {year}
  {2009})%
  \bibAnnoteFile{NoStop}{Krausz_RMP_2009}%
\bibitem{Gertsvolf_JPB_2010}%
  \BibitemOpen
  \bibfield{author}{%
  \bibinfo {author} {\bibfnamefont{M.}~\bibnamefont{Gertsvolf}}, \bibinfo
  {author} {\bibfnamefont{M.}~\bibnamefont{Spanner}}, \bibinfo {author}
  {\bibfnamefont{D.~M.}\ \bibnamefont{Rayner}},\ and\ \bibinfo {author}
  {\bibfnamefont{P.~B.}\ \bibnamefont{Corkum}},\ }%
  \bibfield{journal}{%
  \bibinfo {journal} {Journal of Physics B: Atomic, Molecular and Optical
  Physics}\ }%
  \textbf{\bibinfo {volume} {43}},\ \bibinfo {pages} {131002} (\bibinfo {year}
  {2010})%
  \bibAnnoteFile{NoStop}{Gertsvolf_JPB_2010}%
\bibitem{Mitrofanov_PRL_2011}%
  \BibitemOpen
  \bibfield{author}{%
  \bibinfo {author} {\bibfnamefont{A.~V.}\ \bibnamefont{Mitrofanov}}, \bibinfo
  {author} {\bibfnamefont{A.~J.}\ \bibnamefont{Verhoef}}, \bibinfo {author}
  {\bibfnamefont{E.~E.}\ \bibnamefont{Serebryannikov}}, \bibinfo {author}
  {\bibfnamefont{J.}~\bibnamefont{Lumeau}}, \bibinfo {author}
  {\bibfnamefont{L.}~\bibnamefont{Glebov}}, \bibinfo {author}
  {\bibfnamefont{A.~M.}\ \bibnamefont{Zheltikov}},\ and\ \bibinfo {author}
  {\bibfnamefont{A.}~\bibnamefont{Baltu\ifmmode~\check{s}\else \v{s}\fi{}ka}},\
  }%
  \bibfield{journal}{%
  \Doi{10.1103/PhysRevLett.106.147401}{\bibinfo {journal} {Phys. Rev. Lett.}}\
  }%
  \textbf{\bibinfo {volume} {106}},\ \bibinfo {pages} {147401} (\bibinfo {year}
  {2011})%
  \bibAnnoteFile{NoStop}{Mitrofanov_PRL_2011}%
\bibitem{Ghimire_Nature-Physics_2011}%
  \BibitemOpen
  \bibfield{author}{%
  \bibinfo {author} {\bibfnamefont{S.}~\bibnamefont{Ghimire}}, \bibinfo
  {author} {\bibfnamefont{A.~D.}\ \bibnamefont{DiChiara}}, \bibinfo {author}
  {\bibfnamefont{E.}~\bibnamefont{Sistrunk}}, \bibinfo {author}
  {\bibfnamefont{P.}~\bibnamefont{Agostini}}, \bibinfo {author}
  {\bibfnamefont{L.~F.}\ \bibnamefont{DiMauro}},\ and\ \bibinfo {author}
  {\bibfnamefont{D.~A.}\ \bibnamefont{Reis}},\ }%
  \bibfield{journal}{%
  \Doi{10.1038/NPHYS1847}{\bibinfo {journal} {Nature Physics}}\ }%
  \textbf{\bibinfo {volume} {7}},\ \bibinfo {pages} {138} (\bibinfo {year}
  {2011})%
  \bibAnnoteFile{NoStop}{Ghimire_Nature-Physics_2011}%
\bibitem{Ghimire_PRA_2012}%
  \BibitemOpen
  \bibfield{author}{%
  \bibinfo {author} {\bibfnamefont{S.}~\bibnamefont{Ghimire}}, \bibinfo
  {author} {\bibfnamefont{A.~D.}\ \bibnamefont{DiChiara}}, \bibinfo {author}
  {\bibfnamefont{E.}~\bibnamefont{Sistrunk}}, \bibinfo {author}
  {\bibfnamefont{G.}~\bibnamefont{Ndabashimiye}}, \bibinfo {author}
  {\bibfnamefont{U.~B.}\ \bibnamefont{Szafruga}}, \bibinfo {author}
  {\bibfnamefont{A.}~\bibnamefont{Mohammad}}, \bibinfo {author}
  {\bibfnamefont{P.}~\bibnamefont{Agostini}}, \bibinfo {author}
  {\bibfnamefont{L.~F.}\ \bibnamefont{DiMauro}},\ and\ \bibinfo {author}
  {\bibfnamefont{D.~A.}\ \bibnamefont{Reis}},\ }%
  \bibfield{journal}{%
  \Doi{10.1103/PhysRevA.85.043836}{\bibinfo {journal} {Phys. Rev. A}}\ }%
  \textbf{\bibinfo {volume} {85}},\ \bibinfo {pages} {043836} (\bibinfo {year}
  {2012})%
  \bibAnnoteFile{NoStop}{Ghimire_PRA_2012}%
\bibitem{Ghimire_PRL_2011}%
  \BibitemOpen
  \bibfield{author}{%
  \bibinfo {author} {\bibfnamefont{S.}~\bibnamefont{Ghimire}}, \bibinfo
  {author} {\bibfnamefont{A.~D.}\ \bibnamefont{DiChiara}}, \bibinfo {author}
  {\bibfnamefont{E.}~\bibnamefont{Sistrunk}}, \bibinfo {author}
  {\bibfnamefont{U.~B.}\ \bibnamefont{Szafruga}}, \bibinfo {author}
  {\bibfnamefont{P.}~\bibnamefont{Agostini}}, \bibinfo {author}
  {\bibfnamefont{L.~F.}\ \bibnamefont{DiMauro}},\ and\ \bibinfo {author}
  {\bibfnamefont{D.~A.}\ \bibnamefont{Reis}},\ }%
  \bibfield{journal}{%
  \Doi{10.1103/PhysRevLett.107.167407}{\bibinfo {journal} {Phys. Rev. Lett.}}\
  }%
  \textbf{\bibinfo {volume} {107}},\ \bibinfo {pages} {167407} (\bibinfo {year}
  {2011})%
  \bibAnnoteFile{NoStop}{Ghimire_PRL_2011}%
\bibitem{Baltuska_Nature_2003}%
  \BibitemOpen
  \bibfield{author}{%
  \bibinfo {author} {\bibfnamefont{A.}~\bibnamefont{Baltuska}}, \bibinfo
  {author} {\bibfnamefont{T.}~\bibnamefont{Udem}}, \bibinfo {author}
  {\bibfnamefont{M.}~\bibnamefont{Uiberacker}}, \bibinfo {author}
  {\bibfnamefont{M.}~\bibnamefont{Hentschel}},\ and\ \bibinfo {author}
  {\bibfnamefont{E.}~\bibnamefont{Goulielmakis}},\ }%
  \bibfield{journal}{%
  \Doi{10.1038/nature01414}{\bibinfo {journal} {Nature}}\ }%
  \textbf{\bibinfo {volume} {421}},\ \bibinfo {pages} {611} (\bibinfo {year}
  {2003})%
  \bibAnnoteFile{NoStop}{Baltuska_Nature_2003}%
\bibitem{Schiffrin_Nature_2013}%
  \BibitemOpen
  \bibfield{author}{%
  \bibinfo {author} {\bibfnamefont{A.}~\bibnamefont{Schiffrin}}, \bibinfo
  {author} {\bibfnamefont{T.}~\bibnamefont{Paasch-Colberg}}, \bibinfo {author}
  {\bibfnamefont{N.}~\bibnamefont{Karpowicz}}, \bibinfo {author}
  {\bibfnamefont{V.}~\bibnamefont{Apalkov}}, \bibinfo {author}
  {\bibfnamefont{D.}~\bibnamefont{Gerster}}, \bibinfo {author}
  {\bibfnamefont{S.}~\bibnamefont{Muhlbrandt}}, \bibinfo {author}
  {\bibfnamefont{M.}~\bibnamefont{Korbman}}, \bibinfo {author}
  {\bibfnamefont{J.}~\bibnamefont{Reichert}}, \bibinfo {author}
  {\bibfnamefont{M.}~\bibnamefont{Schultze}}, \bibinfo {author}
  {\bibfnamefont{S.}~\bibnamefont{Holzner}}, \bibinfo {author}
  {\bibfnamefont{J.~V.}\ \bibnamefont{Barth}}, \bibinfo {author}
  {\bibfnamefont{R.}~\bibnamefont{Kienberger}}, \bibinfo {author}
  {\bibfnamefont{R.}~\bibnamefont{Ernstorfer}}, \bibinfo {author}
  {\bibfnamefont{V.~S.}\ \bibnamefont{Yakovlev}}, \bibinfo {author}
  {\bibfnamefont{M.~I.}\ \bibnamefont{Stockman}},\ and\ \bibinfo {author}
  {\bibfnamefont{F.}~\bibnamefont{Krausz}},\ }%
  \bibfield{journal}{%
  \Doi{10.1038/nature11567}{\bibinfo {journal} {Nature}}\ }%
  \textbf{\bibinfo {volume} {493}},\ \bibinfo {pages} {70} (\bibinfo {year}
  {2013})%
  \bibAnnoteFile{NoStop}{Schiffrin_Nature_2013}%
\bibitem{Wannier_PR_1960}%
  \BibitemOpen
  \bibfield{author}{%
  \bibinfo {author} {\bibfnamefont{G.~H.}\ \bibnamefont{Wannier}},\ }%
  \bibfield{journal}{%
  \Doi{10.1103/PhysRev.117.432}{\bibinfo {journal} {Phys. Rev.}}\ }%
  \textbf{\bibinfo {volume} {117}},\ \bibinfo {pages} {432} (\bibinfo {year}
  {1960})%
  \bibAnnoteFile{NoStop}{Wannier_PR_1960}%
\bibitem{Durach_PRL_2011}%
  \BibitemOpen
  \bibfield{author}{%
  \bibinfo {author} {\bibfnamefont{M.}~\bibnamefont{Durach}}, \bibinfo {author}
  {\bibfnamefont{A.}~\bibnamefont{Rusina}}, \bibinfo {author}
  {\bibfnamefont{M.~F.}\ \bibnamefont{Kling}},\ and\ \bibinfo {author}
  {\bibfnamefont{M.~I.}\ \bibnamefont{Stockman}},\ }%
  \bibfield{journal}{%
  \Doi{10.1103/PhysRevLett.107.086602}{\bibinfo {journal} {Phys. Rev. Lett.}}\
  }%
  \textbf{\bibinfo {volume} {107}},\ \bibinfo {pages} {086602} (\bibinfo {year}
  {2011})%
  \bibAnnoteFile{NoStop}{Durach_PRL_2011}%
\bibitem{Keldysh_JETP_1958}%
  \BibitemOpen
  \bibfield{author}{%
  \bibinfo {author} {\bibfnamefont{L.~V.}\ \bibnamefont{Keldysh}},\ }%
  \bibfield{journal}{%
  \bibinfo {journal} {Soviet Phys.\ JETP}\ }%
  \textbf{\bibinfo {volume} {7}},\ \bibinfo {pages} {788} (\bibinfo {year}
  {1958})%
  \bibAnnoteFile{NoStop}{Keldysh_JETP_1958}%
\bibitem{Keldysh_JETP_1965}%
  \BibitemOpen
  \bibfield{author}{%
  \bibinfo {author} {\bibfnamefont{L.~V.}\ \bibnamefont{Keldysh}},\ }%
  \bibfield{journal}{%
  \bibinfo {journal} {Soviet Phys.\ JETP}\ }%
  \textbf{\bibinfo {volume} {20}},\ \bibinfo {pages} {1307} (\bibinfo {year}
  {1965})%
  \bibAnnoteFile{NoStop}{Keldysh_JETP_1965}%
\bibitem{Wegener_2004}%
  \BibitemOpen
  \bibfield{author}{%
  \bibinfo {author} {\bibfnamefont{M.}~\bibnamefont{Wegener}},\ }%
  \emph{\bibinfo {title} {Extreme Nonlinear Optics: An Introduction}},\
  Advanced Texts in Physics\ (\bibinfo {publisher} {Springer},\ \bibinfo {year}
  {2004})\ ISBN \bibinfo {isbn} {9783540222910}%
  \bibAnnoteFile{NoStop}{Wegener_2004}%
\bibitem{Zener_PRSLA_1934}%
  \BibitemOpen
  \bibfield{author}{%
  \bibinfo {author} {\bibfnamefont{C.}~\bibnamefont{Zener}},\ }%
  \bibfield{journal}{%
  \bibinfo {journal} {Proc. R. Soc. London, Ser. A}\ }%
  \textbf{\bibinfo {volume} {145}},\ \bibinfo {pages} {523} (\bibinfo {year}
  {1934})%
  \bibAnnoteFile{NoStop}{Zener_PRSLA_1934}%
\bibitem{Yakovlev_CP_2012}%
  \BibitemOpen
  \bibfield{author}{%
  \bibinfo {author} {\bibfnamefont{V.~S.}\ \bibnamefont{Yakovlev}}, \bibinfo
  {author} {\bibfnamefont{M.}~\bibnamefont{Korbman}},\ and\ \bibinfo {author}
  {\bibfnamefont{A.}~\bibnamefont{Scrinzi}},\ }%
  \enquote{\bibinfo {title} {{Dressed bound states for attosecond dynamics in
  strong laser fields}},}\  (\bibinfo {year} {2012}),\ \bibinfo {note}
  {accepted for Chem.\ Phys.},\
  \Eprint{http://arxiv.org/abs/1110.6783}{arXiv:1110.6783}%
  \bibAnnoteFile{NoStop}{Yakovlev_CP_2012}%
\bibitem{Bachau_PRB_2006}%
  \BibitemOpen
  \bibfield{author}{%
  \bibinfo {author} {\bibfnamefont{H.}~\bibnamefont{Bachau}}, \bibinfo {author}
  {\bibfnamefont{A.~N.}\ \bibnamefont{Belsky}}, \bibinfo {author}
  {\bibfnamefont{P.}~\bibnamefont{Martin}}, \bibinfo {author}
  {\bibfnamefont{A.~N.}\ \bibnamefont{Vasil'ev}},\ and\ \bibinfo {author}
  {\bibfnamefont{B.~N.}\ \bibnamefont{Yatsenko}},\ }%
  \bibfield{journal}{%
  \Doi{10.1103/PhysRevB.74.235215}{\bibinfo {journal} {Phys. Rev. B}}\ }%
  \textbf{\bibinfo {volume} {74}},\ \bibinfo {pages} {235215} (\bibinfo {year}
  {2006})%
  \bibAnnoteFile{NoStop}{Bachau_PRB_2006}%
\bibitem{Blount_1962}%
  \BibitemOpen
  \bibfield{author}{%
  \bibinfo {author} {\bibfnamefont{E.~I.}\ \bibnamefont{Blount}},\ }%
  in\ \Doi{10.1016/S0081-1947(08)60459-2}{\emph{\bibinfo {booktitle} {Solid
  State Physics}}},\ Vol.~\bibinfo {volume} {13},\ \bibinfo {editor} {edited
  by\ \bibinfo {editor} {\bibfnamefont{F.}~\bibnamefont{Seitz}}\ and\ \bibinfo
  {editor} {\bibfnamefont{D.}~\bibnamefont{Turnbull}}}\ (\bibinfo {publisher}
  {Academic Press},\ \bibinfo {year} {1962})\ pp.\ \bibinfo {pages} {305 --
  373}%
  \bibAnnoteFile{NoStop}{Blount_1962}%
\bibitem{Glutsch_PRB_2004}%
  \BibitemOpen
  \bibfield{author}{%
  \bibinfo {author} {\bibfnamefont{S.}~\bibnamefont{Glutsch}},\ }%
  \bibfield{journal}{%
  \Doi{10.1103/PhysRevB.69.235317}{\bibinfo {journal} {Phys. Rev. B}}\ }%
  \textbf{\bibinfo {volume} {69}},\ \bibinfo {pages} {235317} (\bibinfo {year}
  {2004})%
  \bibAnnoteFile{NoStop}{Glutsch_PRB_2004}%
\bibitem{Golde_PRB_2008}%
  \BibitemOpen
  \bibfield{author}{%
  \bibinfo {author} {\bibfnamefont{D.}~\bibnamefont{Golde}}, \bibinfo {author}
  {\bibfnamefont{T.}~\bibnamefont{Meier}},\ and\ \bibinfo {author}
  {\bibfnamefont{S.~W.}\ \bibnamefont{Koch}},\ }%
  \bibfield{journal}{%
  \Doi{10.1103/PhysRevB.77.075330}{\bibinfo {journal} {Phys. Rev. B}}\ }%
  \textbf{\bibinfo {volume} {77}},\ \bibinfo {pages} {075330} (\bibinfo {year}
  {2008})%
  \bibAnnoteFile{NoStop}{Golde_PRB_2008}%
\bibitem{Kazansky_PRL_2009}%
  \BibitemOpen
  \bibfield{author}{%
  \bibinfo {author} {\bibfnamefont{A.~K.}\ \bibnamefont{Kazansky}}\ and\
  \bibinfo {author} {\bibfnamefont{P.~M.}\ \bibnamefont{Echenique}},\ }%
  \bibfield{journal}{%
  \Doi{10.1103/PhysRevLett.102.177401}{\bibinfo {journal} {Phys. Rev. Lett.}}\
  }%
  \textbf{\bibinfo {volume} {102}},\ \bibinfo {pages} {177401} (\bibinfo {year}
  {2009})%
  \bibAnnoteFile{NoStop}{Kazansky_PRL_2009}%
\bibitem{Brabec_2008}%
  \BibitemOpen
  \bibfield{author}{%
  \bibinfo {author} {\bibfnamefont{T.}~\bibnamefont{Brabec}},\ }%
  \emph{\bibinfo {title} {Strong Field Laser Physics}},\ Springer Series in
  Optical Sciences\ (\bibinfo {publisher} {Springer},\ \bibinfo {year} {2008})\
  ISBN \bibinfo {isbn} {9780387400778}%
  \bibAnnoteFile{NoStop}{Brabec_2008}%
\bibitem{Laughlin_PRB_1980}%
  \BibitemOpen
  \bibfield{author}{%
  \bibinfo {author} {\bibfnamefont{R.~B.}\ \bibnamefont{Laughlin}},\ }%
  \bibfield{journal}{%
  \Doi{10.1103/PhysRevB.22.3021}{\bibinfo {journal} {Phys. Rev. B}}\ }%
  \textbf{\bibinfo {volume} {22}},\ \bibinfo {pages} {3021} (\bibinfo {year}
  {1980})%
  \bibAnnoteFile{NoStop}{Laughlin_PRB_1980}%
\bibitem{Malitson_JOSA_1965}%
  \BibitemOpen
  \bibfield{author}{%
  \bibinfo {author} {\bibfnamefont{I.~H.}\ \bibnamefont{Malitson}},\ }%
  \bibfield{journal}{%
  \Doi{10.1364/JOSA.55.001205}{\bibinfo {journal} {J. Opt. Soc. Am.}}\ }%
  \textbf{\bibinfo {volume} {55}},\ \bibinfo {pages} {1205} (\bibinfo {year}
  {1965})%
  \bibAnnoteFile{NoStop}{Malitson_JOSA_1965}%
\bibitem{Chui_1992}%
  \BibitemOpen
  \bibfield{author}{%
  \bibinfo {author} {\bibfnamefont{C.~K.}\ \bibnamefont{Chui}},\ }%
  \emph{\bibinfo {title} {An introduction to wavelets}}\ (\bibinfo {publisher}
  {Academic Press Professional, Inc.},\ \bibinfo {address} {San Diego, CA,
  USA},\ \bibinfo {year} {1992})\ ISBN \bibinfo {isbn} {0-12-174584-8}%
  \bibAnnoteFile{NoStop}{Chui_1992}%
\bibitem{Schneider_PRL_1976}%
  \BibitemOpen
  \bibfield{author}{%
  \bibinfo {author} {\bibfnamefont{P.~M.}\ \bibnamefont{Schneider}}\ and\
  \bibinfo {author} {\bibfnamefont{W.~B.}\ \bibnamefont{Fowler}},\ }%
  \bibfield{journal}{%
  \Doi{10.1103/PhysRevLett.36.425}{\bibinfo {journal} {Phys. Rev. Lett.}}\ }%
  \textbf{\bibinfo {volume} {36}},\ \bibinfo {pages} {425} (\bibinfo {year}
  {1976})%
  \bibAnnoteFile{NoStop}{Schneider_PRL_1976}%
\bibitem{Chelikowsky_PRB_1977}%
  \BibitemOpen
  \bibfield{author}{%
  \bibinfo {author} {\bibfnamefont{J.~R.}\ \bibnamefont{Chelikowsky}}\ and\
  \bibinfo {author} {\bibfnamefont{M.}~\bibnamefont{Schl\"uter}},\ }%
  \bibfield{journal}{%
  \Doi{10.1103/PhysRevB.15.4020}{\bibinfo {journal} {Phys. Rev. B}}\ }%
  \textbf{\bibinfo {volume} {15}},\ \bibinfo {pages} {4020} (\bibinfo {year}
  {1977})%
  \bibAnnoteFile{NoStop}{Chelikowsky_PRB_1977}%
\bibitem{Bloch_ZPA_1929}%
  \BibitemOpen
  \bibfield{author}{%
  \bibinfo {author} {\bibfnamefont{F.}~\bibnamefont{Bloch}},\ }%
  \bibfield{journal}{%
  \Doi{10.1007/BF01339455}{\bibinfo {journal} {Zeitschrift f\"ur Physik A
  Hadrons and Nuclei}}\ }%
  \textbf{\bibinfo {volume} {52}},\ \bibinfo {pages} {555} (\bibinfo {year}
  {1929})%
  \bibAnnoteFile{NoStop}{Bloch_ZPA_1929}%
\bibitem{Delone_2000}%
  \BibitemOpen
  \bibfield{author}{%
  \bibinfo {author} {\bibfnamefont{N.}~\bibnamefont{Delone}}\ and\ \bibinfo
  {author} {\bibfnamefont{V.}~\bibnamefont{Kraǐnov}},\ }%
  \emph{\bibinfo {title} {Multiphoton Processes in Atoms}},\ Springer Series on
  Atoms + Plasmas\ (\bibinfo {publisher} {Springer},\ \bibinfo {year} {2000})\
  ISBN \bibinfo {isbn} {9783540646150}%
  \bibAnnoteFile{NoStop}{Delone_2000}%
\bibitem{Kane_JPCS_1960}%
  \BibitemOpen
  \bibfield{author}{%
  \bibinfo {author} {\bibfnamefont{E.~O.}\ \bibnamefont{Kane}},\ }%
  \bibfield{journal}{%
  \Doi{10.1016/0022-3697(60)90035-4}{\bibinfo {journal} {J.\ of Phys. and
  Chem.\ of Solids}}\ }%
  \textbf{\bibinfo {volume} {12}},\ \bibinfo {pages} {181 } (\bibinfo {year}
  {1960})%
  \bibAnnoteFile{NoStop}{Kane_JPCS_1960}%
\end{thebibliography}%


\end{document}